\documentclass[10pt,linenumbers]{iopart}
  
\usepackage[utf8]{inputenc}

\usepackage{amsfonts,amssymb}
\usepackage{bm}

\usepackage{url}
\usepackage{graphicx}
\usepackage{setspace}

\usepackage{microtype}
\usepackage[colorlinks=true,linkcolor=blue,urlcolor=blue,citecolor=blue]{hyperref}

\def\be{\begin{eqnarray} &&} 
 
\def\ee{\end{eqnarray}}

\def\lf{\left}
\def\rg{\right}

\graphicspath{{.}{plots/}}

\date{\today}

\begin{document}

\title{Exposing Novel Quark and Gluon Effects in Nuclei}

\author{I.~C.~Clo\"et}
\address{Physics Division, Argonne National Laboratory, Lemont, IL 60439, USA}

\author{R.~Dupr\'{e}}
\address{Institut de Physique Nucl\'eaire, CNRS-IN2P3, Univ. Paris-Sud, Universit\'e Paris-Saclay, 91406 Orsay Cedex, France}

\author{S.~Riordan}
\address{Physics Division, Argonne National Laboratory, Lemont, IL 60439, USA}
\ead{sriordan@anl.gov (Corresponding author)}

\author{W.~Armstrong}
\address{Physics Division, Argonne National Laboratory, Lemont, IL 60439, USA}

\author{J.~Arrington}
\address{Physics Division, Argonne National Laboratory, Lemont, IL 60439, USA}

\author{W.~Cosyn}
\address{Department of Physics and Astronomy, Ghent University, Proeftuinstraat 86, B9000 Ghent, Belgium}

\author{N.~Fomin}
\address{University of Tennessee, Knoxville, TN 37996, USA}

\author{A.~Freese}
\address{Physics Division, Argonne National Laboratory, Lemont, IL 60439, USA}

\author{S.~Fucini}
\address{Perugia University and INFN, Perugia Section, via A. Pascoli snc I-06123 Perugia, Italy}

\author{D.~Gaskell}
\address{Thomas Jefferson National Accelerator Facility, Newport News, VA 23606, USA}

\author{C.~E.~Keppel}
\address{Thomas Jefferson National Accelerator Facility, Newport News, VA 23606, USA}

\author{G.~A.~Miller}
\address{Department of Physics, University of Washington, Seattle, WA 98195-1560}

\author{E.~Pace}
\address{Universit\`a di Roma ``Tor Vergata'' and INFN, Sezione di  Roma Tor Vergata, 00133 Rome, Italy}

\author{S.~Platchkov}
\address{IRFU, CEA, Université Paris-Saclay, 91191 Gif-sur-Yvette, France}

\author{P.~E.~Reimer}
\address{Physics Division, Argonne National Laboratory, Lemont, IL 60439, USA}

\author{S.~Scopetta}
\address{Università di Perugia and INFN, Sezione di Perugia, via A. Pascoli snc I-06123 Perugia, Italy}

\author{A.~W.~Thomas}
\address{ARC Centre of Excellence for Particle Physics at the Terascale and CSSM, Department of Physics,
University of Adelaide, Adelaide SA 5005, Australia}

\author{P.~Zurita}
\address{Physics Department, Brookhaven National Laboratory, Upton, NY 11973, USA}


\begin{abstract}
The fundamental theory of the strong interaction -- quantum chromodynamics (QCD) -- provides the foundational framework with which to describe and understand the key properties of atomic nuclei. A deep understanding of the explicit role of quarks and gluons in nuclei remains elusive however, as these effects have thus far been well-disguised by confinement effects in QCD which are encapsulated by a successful description in terms of effective hadronic degrees of freedom. The observation of the EMC effect has provided an enduring indication for explicit QCD effects in nuclei, and points to the medium modification of the bound protons and neutrons in the nuclear medium. Understanding the EMC effect is a major challenge for modern nuclear physics, and several key questions remain, such as understanding its flavor, spin, and momentum dependence. This manuscript provides a contemporary snapshot of our understanding of the role of QCD in nuclei and outlines possible pathways in experiment and theory that will help deepen our understanding of nuclei in the context of QCD.
\end{abstract}

\submitto{\jpg}
\maketitle
\ioptwocol

\begin{spacing}{0.9}
\tableofcontents
\end{spacing}

\section{Introduction}
How does the nucleus arise from quantum chromodynamics (QCD)? Answering this question is a key challenge for modern science. Traditional nuclear physics regards the nucleus as being composed of bound nucleons and mesons. This picture has had significant success in describing the properties of nuclei across the chart of nuclides. However, the fundamental theory of the strong interaction is QCD, where quarks and gluons are the elementary degrees of freedom. This means that it is unlikely that the nucleon-meson based approaches can remain valid or contain the correct degrees of freedom for all processes at all energy scales. Clearly identifying these scales and processes is key to exposing the role of quarks and gluons in nuclei and thereby developing an understanding of how nuclei emerge within QCD.

Deep inelastic scattering experiments have long suggested that a nucleon-meson based picture of the nucleus is incomplete, however to gain a more detailed understanding of the quark-gluon structure of nuclei a broad program in experiment and theory must be developed. This includes novel measurements of nuclear structure with high energy leptonic probes with inclusive, semi-inclusive and exclusive final states, Drell-Yan processes with different incident hadrons on a variety of nuclei, a rigorous development of theoretical frameworks and modeling, and careful constraint and understanding of systematic effects.

Determining answers to several key questions is now possible however. These answers would greatly expand our knowledge of how the nucleus influences the quarks and gluons within bound nucleons. Such effects are called medium modifications. This includes the nature of the isovector nuclear forces and their impact on the various parton distribution functions (PDFs) of nuclei, nuclear spin-dependent PDFs, the relation between the momenta of the bound internal quarks and the hadronic constituents, and the full femtoscopic imaging of the nucleus.  Coupled with these studies is the need for a rigorous formalism and a better understanding of the systematic effects in processes such as hadronization and in the extraction of nuclear longitudinal structure functions. 

The goal of the article is to provide a status of the current experimental and theoretical understanding of the role of QCD in nuclei and to provide a road map containing a key set of questions for the next era of measurements and calculations. These new directions in experiment and theory will cover needed information for the latest nuclear parton distribution functions, programs which will study the spin and isospin-dependence of medium modification, better constrain both valence and sea distributions, and ultimately achieve a more complete tomography of the structure of nuclei. 

The structure of the article is as follows: Sec.~\ref{sec:status} provides a summary of the current status of the EMC effect; Sec.~\ref{sec:nPDFs} discusses nuclear PDFs; Sec.~\ref{sec:directions} outlines some future directions for unpolarized lepton scattering measurements including measurements of short-ranged correlations; Secs.~\ref{sec:ivemc} and \ref{sec:pemc} discuss the isovector and polarized EMC effects; Sec.~\ref{sec:DY} discusses Drell-Yan measurements; Sec.~\ref{sec:light} looks at opportunities with light nuclei such as the deuteron; Sec.~\ref{sec:lf} discusses light-front methods; Sec.~\ref{sec:tagged} studies tagged reactions; Sec.~\ref{sec:GPDs} discusses nuclear GPDs;  nuclear effects in longitudinal structure functions in Sec.~\ref{sec:long}; systematic effects are studied in Sec.~\ref{sec:systematics}, and finally a summary and outlook is given in Sec.~\ref{sec:conclusion}.

\section{Status of the EMC effect\label{sec:status}}
One of the best tools available for studying the internal structure of nucleons is the deep inelastic
scattering (DIS) process, where a charged lepton scatters inelastically with a large four-momentum
transfer, $q^\mu$, represented by the Lorentz invariant $Q^2 = -q^\mu q_\mu \gg 1-2~\mathrm{GeV}^2$.
The process is characterized by a scaling variable called Bjorken-$x$, defined by $x = Q^2/2M \nu$, 
where $M$ is the mass of the target and $\nu$ is the energy transferred from the lepton in the laboratory frame. Bjorken-$x$ is associated with the light-cone momentum fraction carried by the struck parton in the limit of the infinite-momentum frame. To leading order the unpolarized electromagnetic DIS cross-section for a spin-1/2 target such as the proton can be written as~\cite{PhysRevD.98.030001,Taylor:1991ew}
\begin{equation}
\frac{d^2 \sigma}{dx dy} = \frac{4 \pi \alpha^2}{x y Q^2} \left[ (1-y)F_2(x, Q^2) + y^2 x F_1(x, Q^2) \right],
\end{equation}
where $y = \nu/E$ with $E$ the incident lepton energy, $\alpha$ is the electromagnetic fine structure constant, and $F_2$
and $F_1$ are the structure functions of the target to be determined by experiment. These structure functions can be reduced by the Callan-Gross relation $F_2 - 2xF_1 = F_\mathrm{L} \approx 0$ for non-interacting point-like spin-1/2 particles with
\begin{equation}
F_2(x,Q^2) = x \sum_{q} e_q^2 \left[q(x,Q^2) + \bar{q}(x,Q^2)\right],
\end{equation}
where the sum $q$ is over all quark flavors,  $e_q$ is the electromagnetic charge of the quark, and the functions $q(x,Q^2)$ and $\bar{q}(x,Q^2)$ are the PDFs which are universal and can describe other processes such as Drell-Yan scattering.  The ratio of $F_L$ to $F_T$, the transverse structure function, $F_T$,  $R = F_L/F_T$ is of important interest for small $x$ and small $Q^2$ where the contributions can be sizeable from photon-gluon fusion and higher-twist corrections respectively.  This has not been as well studied for nuclear targets and often represents an open systematic uncertainty.  Analogous distributions exist for the case of polarized targets. These PDFs can in principle be measured for any bound hadronic state.

\begin{figure}[tbp]
\centering\includegraphics[width=\columnwidth]{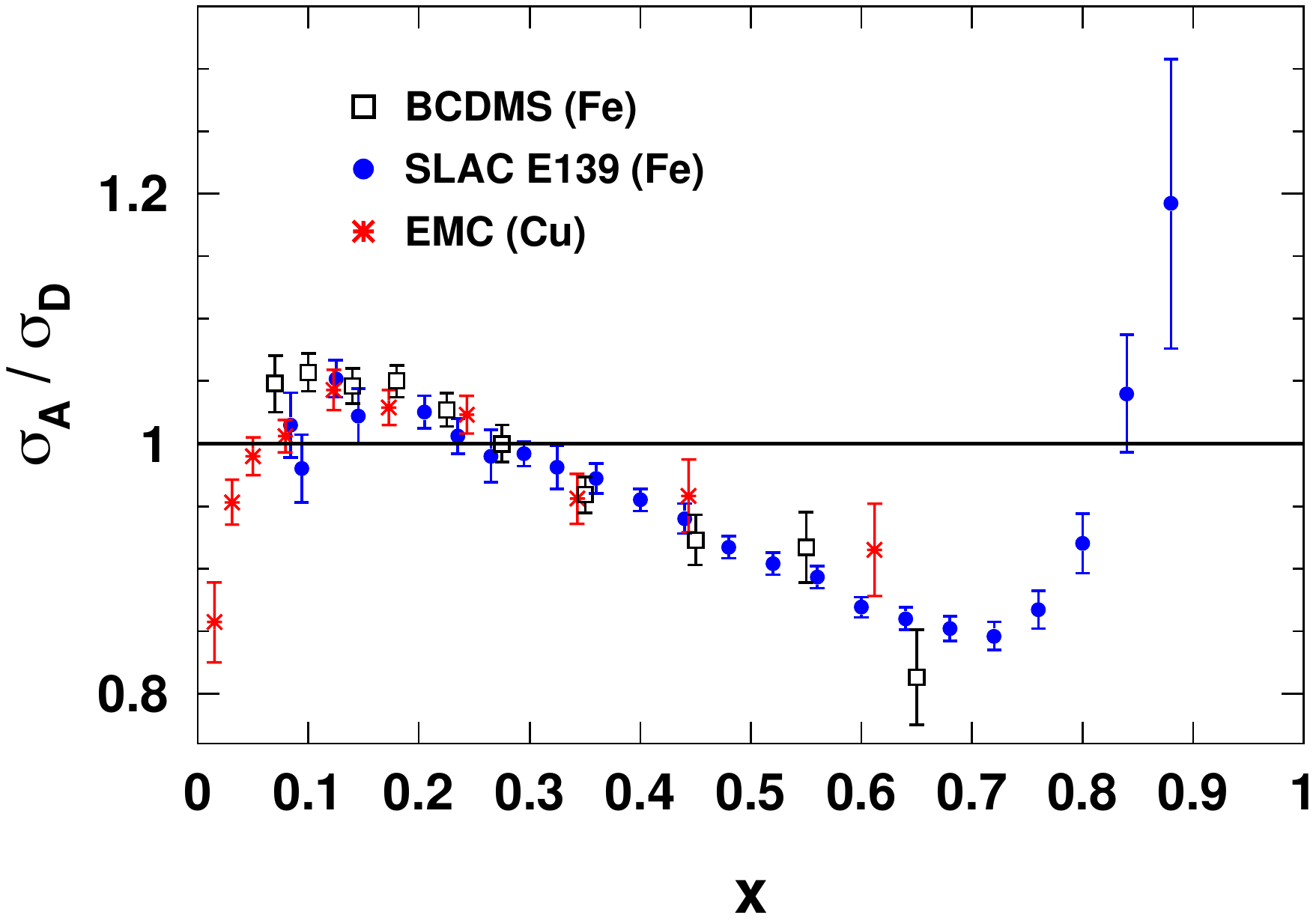}
\caption{EMC effect for iron (BCDMS collaboration~\cite{Benvenuti:1987az} and SLAC E139~\cite{Gomez:1993ri}) and copper (EMC collaboration~\cite{Ashman:1992kv}).
Figure from Ref.~\cite{Guzey:2012yk}.}
\label{fig:emc_iron}
\end{figure}

The EMC effect~\cite{Aubert:1983xm} is the observation that the PDFs for nuclei are different than the incoherent sum over the PDFs of the constituent nucleons. Together with Coulomb sum rule measurments~\cite{Altemus:1980wt,Noble:1980my,Meziani:1984is,Morgenstern:2001jt,Cloet:2015tha}, the EMC effect was among the first evidence that the structure of the nucleons may be different when bound together in a nucleus.\footnote{The term EMC stands for the European Muon Collaboration which originally discoveried this effect.} Since the original discovery in 1983, there has been a large program of measurements at several laboratories, such as CERN, Fermilab, SLAC, DESY, and Jefferson Lab (JLab), aimed at understanding the properties and probing the origin of the nuclear dependence of inelastic structure functions, covered in detail in several reviews~\cite{Frankfurt:1988nt,Arneodo:1992wf,Geesaman:1995yd,Piller:1999wx,Sargsian:2002wc,Malace:2014uea,Hen:2016kwk}. Measurements with high energy muon beams (on the scale of 100~GeV) have provided high-precision data at low to moderate $x$ ($<0.3$), while more modest energy electron facilities (on scale of 10 GeV) have provided the highest precision at large $x$ ($>0.3$). See  Fig.~\ref{fig:emc_iron} for data on iron and copper targets.  The low-to-moderate $x$ regions show interesting shadowing and anti-shadowing behaviors, while the suppression of the per-nucleon cross section for $0.3<x<0.7$ is the hallmark of the EMC effect. For $x > 0.7$ effects from the Fermi motion of the nucleons in the nucleus begin to dominate, which causes the rapid rise in the cross-section ratio in this region.

\begin{figure}[tbp]
\centering\includegraphics[width=\columnwidth]{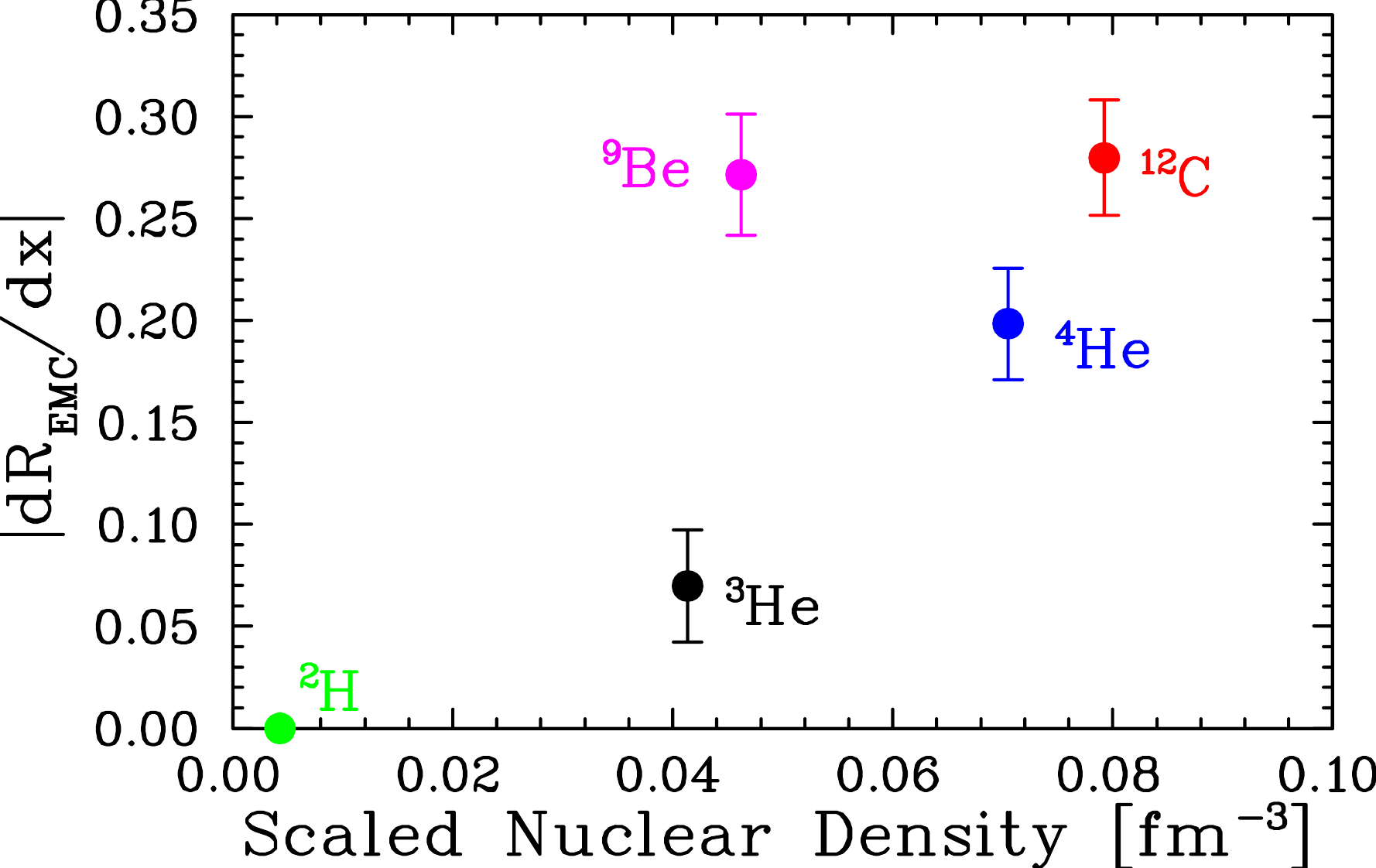}
\caption{Size of the EMC effect, in this case assumed to be the slope $|dR/dx|$ between $x=0.35-0.70$, vs. scaled nuclear density. Figure adapted from Ref.~\cite{Seely:2009gt}.}
\label{fig:emc_jlab_hallc}
\end{figure}

The most comprehensive data sets with high precision at large $x$ come from SLAC E139~\cite{Gomez:1993ri} and JLab E03-103~\cite{Seely:2009gt}. The SLAC experiment
measured the EMC effect for a wide range of nuclei, from $^4$He to Au with good precision up to
$x\approx0.8$.  One of the outcomes of E139 was an investigation of the detailed nuclear dependence of the EMC
effect. It was found that the nuclear dependence of the EMC effect at $x \gtrsim =0.6$ is consistent
with both a logarithmic $A$ dependence or a linear dependence on average nuclear density (often parametrized as an $A^{1/3}$ dependence).

The SLAC and early high energy measurements showed a number of global properties of the EMC effect:
\begin{itemize}
\item{The shape of the EMC effect (shadowing, anti-shadowing, and EMC regions at small, moderate, and large $x$ respectively) is universal and observed in all nuclei.}
\item{The EMC effect displays little $Q^2$ dependence over the full $x$ range.}
\item{At small $x$ and large $x$, the EMC effect grows with $A$, while there is little apparent $A$ dependence in the anti-shadowing region.}
\end{itemize}
Since the first observation of the EMC effect, many theoretical models have been proposed and can be subdivided into two categories.  One includes only ``traditional'' nuclear physics effects, using convolution models with binding effects, detailed models of the nucleon momentum distribution, or pion-exchange contributions. The other category invokes more exotic explanations such as re-scaling of quark distributions in the nuclear environment, contributions of six or nine
quark bags, or modification of the internal structure of the nucleons such as ``nucleon swelling'' or suppression of point-like nucleon configurations. Several reviews give an overview of models of the EMC effect~\cite{Geesaman:1995yd, Piller:1999wx, Malace:2014uea, Norton:2003cb,  Hen:2013oha}.

\begin{figure}[tbp]
\centering\includegraphics[width=\columnwidth]{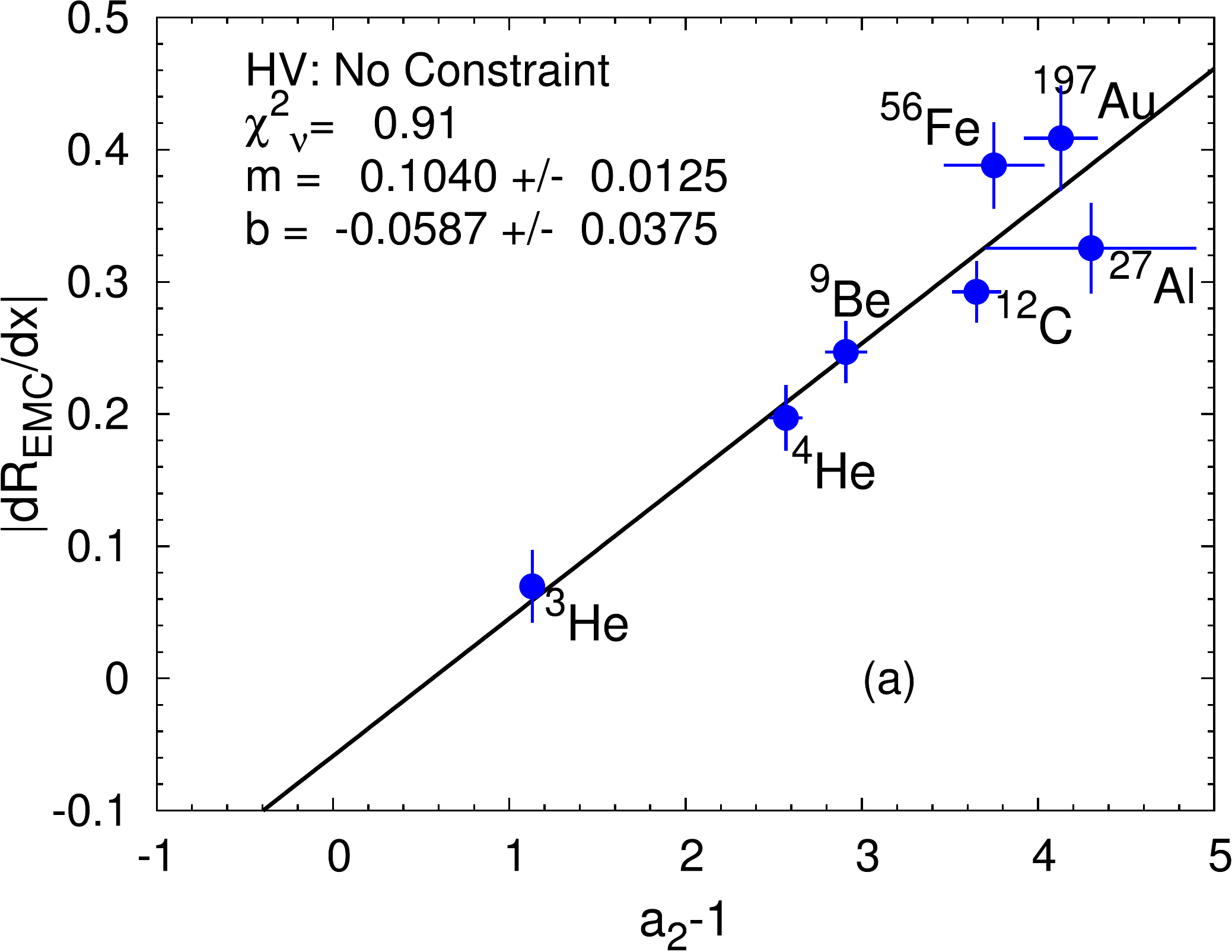}
\caption{Size of the EMC effect plotted vs. the SRC ratio $a_2$. Data are from JLab and SLAC. Figure
from Ref.~\cite{Arrington:2012ax}.}
\label{fig:emc_src_bff}
\end{figure}

\subsection{EMC effect results from Jefferson Lab and the EMC-SRC correspondence}

The primary goal of Jefferson Lab E03-103 was to augment the results obtained from SLAC E139 by taking
advantage of improved target technologies to provide higher precision data for $^4$He, and the first
measurement of the EMC effect from $^3$He at large $x$.  Additional light (Be and C) and heavy (Cu and Au)
targets were also used to provide improved precision at large $x$.

The EMC effect was described using the ratio $R$ of the measured nucleus cross section and deuterium
normalized to the relative number of nucleons.
Since the shape of the EMC effect is universal, the E03-103 analysis assumed the ``size'' of the EMC
effect as the slope of a linear fit of $R$ between $0.35<x<0.7$.  This definition reduces sensitivity to
normalization uncertainties and results in higher sensitivity to the nuclear dependence.

The slope $R$ from E03-103 for $^3$He, $^4$He, Be, and C plotted against scaled nuclear
density is shown in Fig.~\ref{fig:emc_jlab_hallc}.  The nuclear dependence was found to be consistent
with a simple density dependence, with the exception of $^9$Be,  which showed an anomalously larger-than-expected 
EMC effect given its low average nuclear density.  However, if the size of the EMC effect is
governed by the local nuclear density experienced by the quarks in the bound nucleon, rather than the
average density~\cite{Seely:2009gt, PhysRevC.82.054614} a different size of the EMC effect would be predicted.

\begin{figure*}[tbp]
\includegraphics[width=1.0\columnwidth]{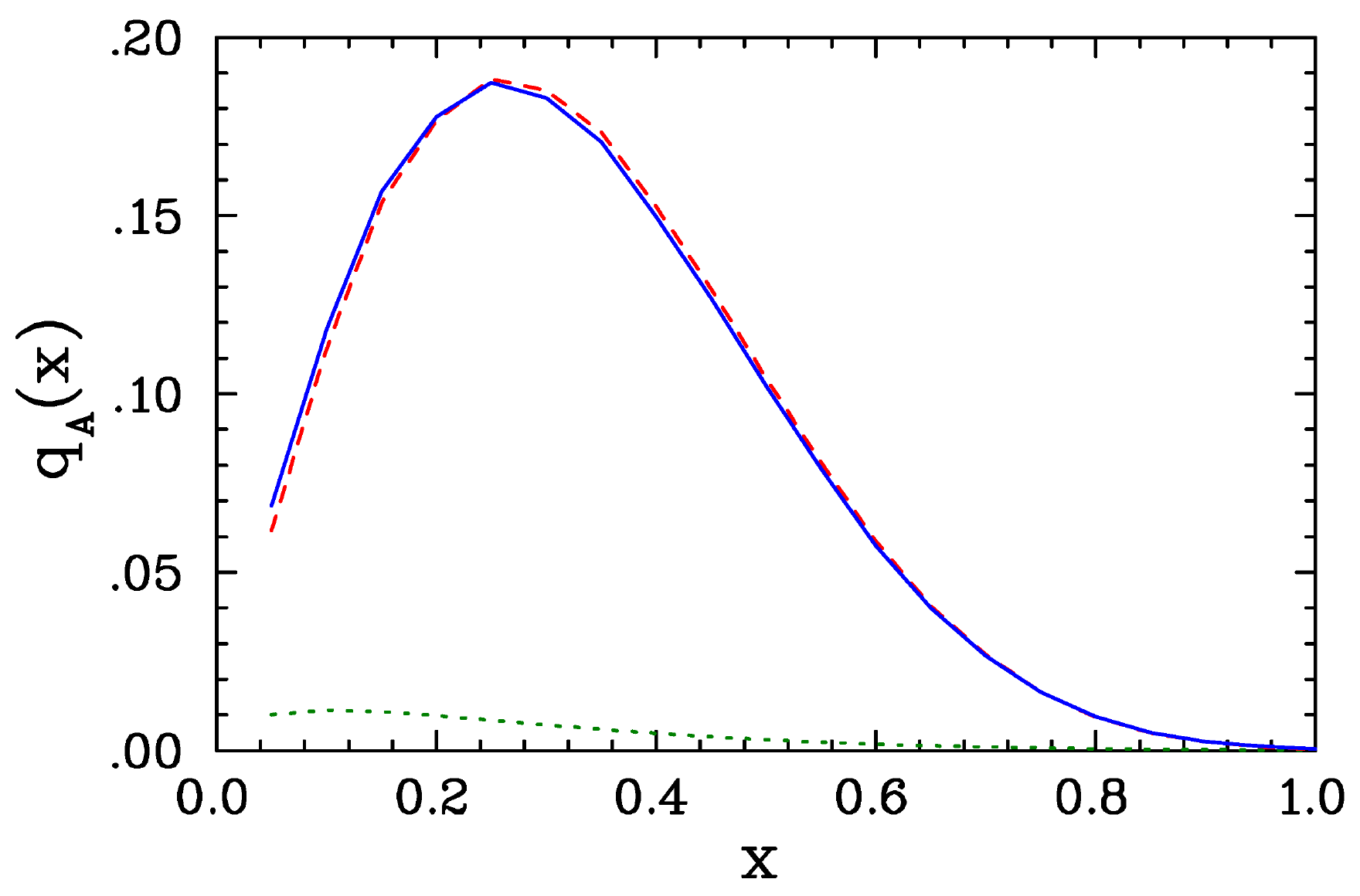} \hfill
\includegraphics[width=1.0\columnwidth]{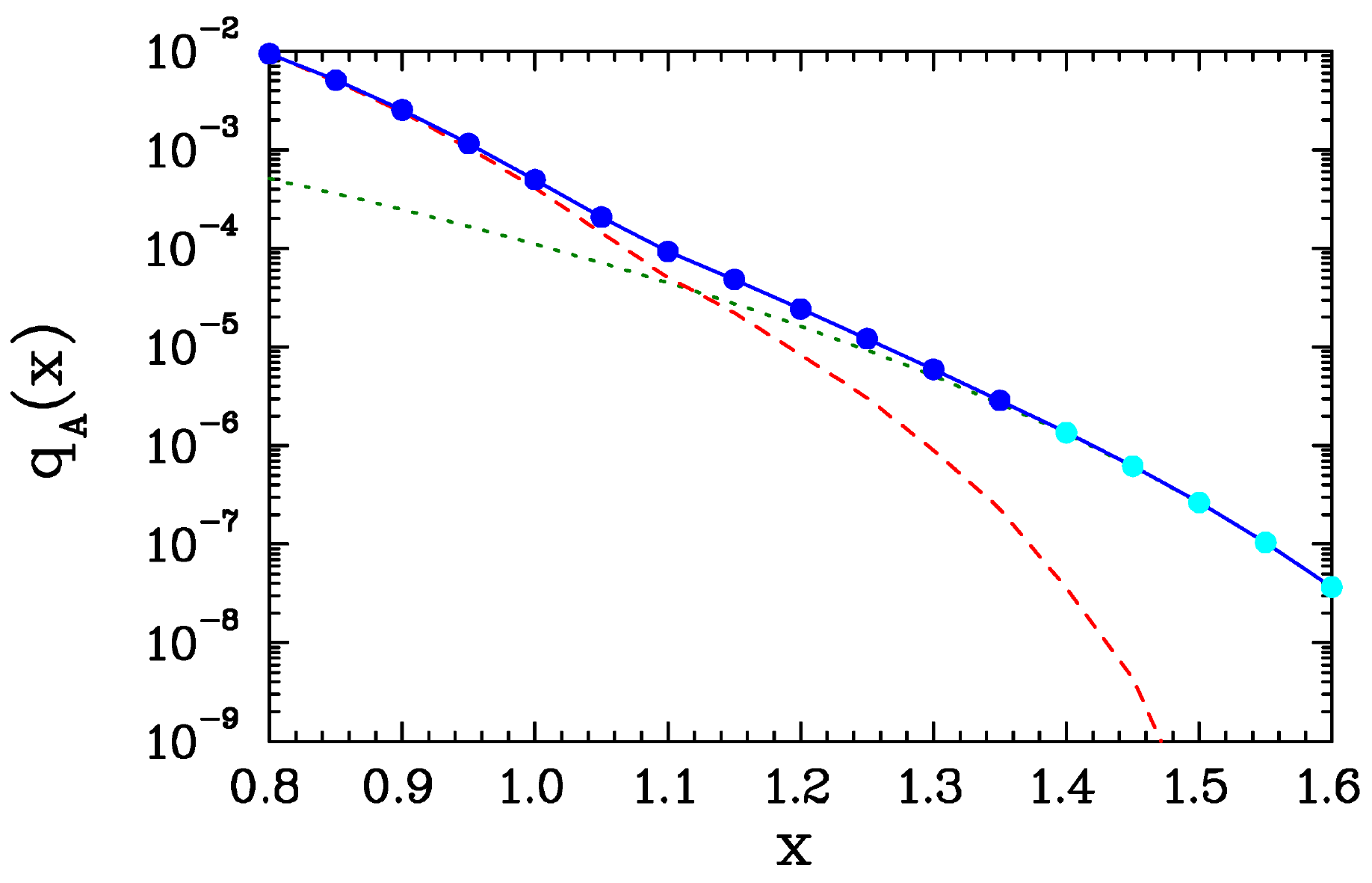}
\caption{This figure is from Ref.~\cite{Arrington:2003qt}, and presents quark distribution calculations for the deuteron, where the dashed line shows 2N components, the dotted line corresponds to a 5\% contribution from 6-quark bags, and the solid blue line shows the sum of the two~\cite{Mulders:1983au}. }
\label{fig:quarkbags}
\end{figure*}

When the EMC effect was compared to the unnormalized ratio $a_2=\sigma_A/\sigma_D$~\cite{Frankfurt:1993sp} of the
inclusive cross sections at $x>1$ (which is sensitive to the ``number of short range correlations''
found in a nucleus) a linear correlation between the two quantities was observed~\cite{Weinstein:2010rt}.
This correspondence was even more compelling when $x>1$ ratios for $^9$Be from JLab experiment E02-019 became
available and the same correlation persisted~\cite{Arrington:2012ax, Hen:2012fm}, Fig.~\ref{fig:emc_src_bff}.

The origin of this correspondence is unclear, whether short-ranged correlations (SRCs) in some way cause the EMC effect, or if the two
phenomena are caused by some common underlying source.  A study was conducted to examine whether
both the EMC effect and SRCs are correlated with other independent variables~\cite{Arrington:2012ax}, such
as average nucleon separation energy, with no clear common origin or factor found.  Separate calculations have been published using effective field theory which show this is a natural consequence of scale separation~\cite{PhysRevLett.119.262502}.

A subtlety exists in correlation studies between $x>1$ inclusive cross section ratios and the EMC effect. The former represent the relative number of high-momentum nucleons found in a nucleus of mass number $A$ to $A=2$, and not the relative number of nucleon-nucleon pairs.  This means that a causal relationship between high-virtuality nucleons and the EMC effect can potentially represent a different scenario than one where the high-density configurations (SRC pairs) are the culprit.  

\subsection{Quarks at $x>1$}
Quasielastic kinematics at $x>1$ probe moving nucleons, but at larger $Q^2$, the inelastic contribution begins to dominate, giving access to quark distributions. Existing JLab data have a limited range in $x$ and require the application of the of so-called "target-mass corrections" (TMCs) to extract the $Q^2 \to \infty$ structure function limit.  This analysis was done for the E02-019 data, showing that the data are approaching the scaling regime~\cite{Fomin:2010ei}.  Here potential 6-quark configurations could result in a noticeable enhancement of the measured cross sections at $x>1$, and be related back to the EMC region where the enhancement is harder to determine.  Fig.~\ref{fig:quarkbags} from Ref.~\cite{Arrington:2003qt} shows the potential effect on the quark distributions in the EMC and the $x>1$ region.


\section{Nuclear PDFs\label{sec:nPDFs}}
Nuclear parton distributions functions (nPDFs) are the first step in understanding the behavior of nuclear matter at the elementary particle level. Moreover they play a crucial role in the determination of the free proton PDFs, as nuclear targets are routinely used for separating the different partonic flavors in PDFs fits. Notwithstanding their importance, nPDFs are not as well known as the free proton case, primarily due to two factors. First, despite the phenomenal success of HERA in determining the proton structure, no electron-nucleus collider has ever existed. Second, only a few nuclei have been studied in detail and the data span a limited region of the kinematic space, to the point that the only constrained nuclear distributions are the distributions quarks in the mid-$x$ region.  For regions where data are not readily available, extrapolations for $x < 10^{-3}$ employ the fulfillment of the charge and momentum sum rules and, at mid to high-$x$, the sea quarks and gluon densities are often determined by ad-hoc assumptions during the fitting procedure, rather than from actual constraints from the data.  The parametrizations of the nuclear effects in the neutron frequently employ the assumption of isospin symmetry, which is difficult to validate given the nearly isoscalar nature of most available nuclei.

Less than about one third of the data used in nPDF fits come from heavy nuclei which complicates the possibility of truly separating the nuclear modifications for each flavor. A possible path for flavor separation would be using charged current (CC) data from DIS~\cite{Tzanov:2005kr,Onengut:2005kv,Tzanov:2009zz}, which is available for iron and lead, where the cross-section depends on different combinations of the PDFs than the neutral current (NC) processes. It has been also suggested that nuclear effects might not be universal and therefore making a truly global fit of the nPDFs would not be possible. Up to now and within experimental uncertainties, NLO fits including NC and CC data have not shown visible tension. Unfortunately these fixed target experiments cover a very limited region of the kinematic space and are lacking in precision, not allowing yet for a conclusive answer.

The unexplored low-$x$ region, dominated by the gluon density, opens the possibility of finding new non-linear phenomena such saturation, and puts to test the applicability range of the linear regime. The other extreme of the kinematic space, the high-$x$ region, is of particular interest as there appears the first measured sign of nuclear effects in high energy collisions: the EMC effect. Moreover, for beyond the Standard Model searches at the LHC, rare high-$x$ gluon initiated events could be enhanced. However the nuclear gluon is difficult to access at high-$x$, and great care has to be put in its determination. Despite the lack of data, the determination of nPDFs in all the kinematic space is of crucial relevance and thus the target of several efforts. Given the fact that sea quarks and gluon densities are tied through the DGLAP evolution equations, studying processes sensitive to either sea quarks or gluons has an impact on our knowledge of nPDFs.    

\subsection{Accessing the Nuclear Gluon at High-$x$}

One of the observables in which initial state gluons can account for most of the cross-section are jets. Unfortunately the LHC measurements in jets in $p+\mathrm{Pb}$ collisions by the ATLAS collaboration~\cite{ATLAS:2014cpa} have been integrated over in the $0-90\%$ centrality bin instead of using minimum bias data, rendering the quantity difficult for comparison to collinear factorized pQCD predictions. However, the data of the di-jets measured by the CMS collaboration were published~\cite{Chatrchyan:2014hqa} and further included in the latest NLO nPDF analysis of EPPS16~\cite{Eskola:2016oht}. There it was shown that the di-jets from CMS have a non-negligible impact on the high-$x$ gluon distribution. As the EPPS16 fit comprises about $2000$ data points and allows for more flexible parametrizations, the effect of the di-jet data on constraining the gluon is less effective. 

Nonetheless, jets remain a relevant tool to access the gluon density. In recent works~\cite{PhysRevD.95.094013, PhysRevD.97.114013} it has been shown that (di-~)jets in $e+A$ collisions at a future Electron-Ion Collider (EIC) have the potential to reduce the theoretical uncertainties by an order of magnitude at both low and high-$x$, reaching into the anti-shadowing and EMC effect regions.

A complementary way of accessing the high-$x$ gluon is using the charm quark structure function. This quantity is determined by tagging the charm in the final state and theoretically has its leading order perturbative contribution from the photon-gluon fusion process. In addition, this observable might hold the key to study if there is an intrinsic content of charm in the proton or nucleus or if heavy quarks appear only by radiation from the gluons, though fully disentangling the gluon and intrinsic charm is challenging. The studies of DIS reduced cross-section with simulated EIC data and its impact on the gluon nPDF~\cite{PhysRevD.96.114005} show that the inclusive data could reduce the uncertainty bands up to a factor of~4 at low $x$, while the charm reduced cross-section would have a dramatic effect, diminishing the uncertainties by almost an order of magnitude at high-$x$.   

\subsection{Improvements for nPDFs}

Future programs will play a key role on determining the nPDFs and it is crucial to incorporate lessons learned from prior experiments, publications, and analysis. A simple example is results from HERA where initial data were published in the form of the structure functions $F_{2}$, while in later years and after the conclusion of data taking, new publications were made of the cross-sections. The shift came from the realization that the structure functions are not a fundamental observable if considering the gluon densities and potential saturation effects. This adds uncertainty in $F_{2}$ and no reanalysis under this realization has been performed.  Similarly, assumptions are made about the form of the longitudinal structure functions $F_\mathrm{L}$ and become deeply ingrained in the results.  Corrections 
to account for the non-isoscalarity of some targets make assumptions on the flavor-dependence of the modification in nuclear data analyses and can lead to very different shapes for the EMC effect.  In this light it would be extremely beneficial for the community to publish future results which include the measured cross sections, as well as the extracted structure functions, with and without corrections or assumptions.

\section{Future Directions for Unpolarized Lepton Scattering\label{sec:directions}}

With the observations from JLab's 6~GeV era that the EMC effect may be driven by the local nuclear environment
and the apparent correspondence between the EMC effect and short range correlations, it is worth examining what
further studies with leptons scattering can be performed.
Short-range correlations offer access to understanding a more detailed picture of the high-momentum structure of nuclei as well as possible unique insight into the generation of the EMC effect.  Inclusive experiments aim to measure precision cross section ratios at $x$ above and below unity, with $x > 1$ a region forbidden to the free nucleon.  Exclusive measurements can yield a more detailed picture of the high momentum structure of nuclei, its isospin structure, and many-body correlations.

One clear avenue of exploration is to make inclusive EMC effect region measurements from additional light to medium-light
nuclei where ab-initio nuclear structure calculations are feasible and interesting nuclear cluster
structures may manifest.  One can also leverage the fact that SRCs are dominated by neutron-proton pairs to further explore
the EMC-SRC correspondence by measuring the EMC effect for a range of nuclei with
different neutron to proton ($N/P$) ratio at fixed $A$, and for a range of $A$ at fixed $N/P$ which would
expose an isovector dependence.

Inclusive $x>1$ scattering offers several unique opportunities.  Here, the lepton probes high-momentum nucleons, typically defined as ones with momenta $\ge$300~MeV/c, about a few tens of~MeV greater than the Fermi momentum for a given nucleus of $A>12$.  If the presence of these high-momentum nucleons is the result of SRCs, then the cross sections for $A>2$ will be re-scaled versions of the deuteron cross sections, signaling more SRCs and therefore more high-momentum nucleons in larger nuclei.   The observation of a plateau at $x>1$ in the $A/D$ cross-section ratios supports this picture. 

 Experimentally, the onset of the $x>1$ plateau is observed to occur at lower $x$ values as $Q^2$ increased, with the minimum $Q^2$ value taken to be about 1.4~GeV$^2$ illustrated with data from Ref.~\cite{Egiyan:2003vg}.  Cross section plateaus at $x>1$ were observed in several JLab experiments~\cite{Egiyan:2003vg, Fomin:2011ng} and is proportional to the relative number of high-momentum nucleons in $A$ with respect $D$, including the number of SRC pairs and higher order correlations.   

\begin{figure}[tbp]
\includegraphics[width=\columnwidth,width=55mm, angle=-90]{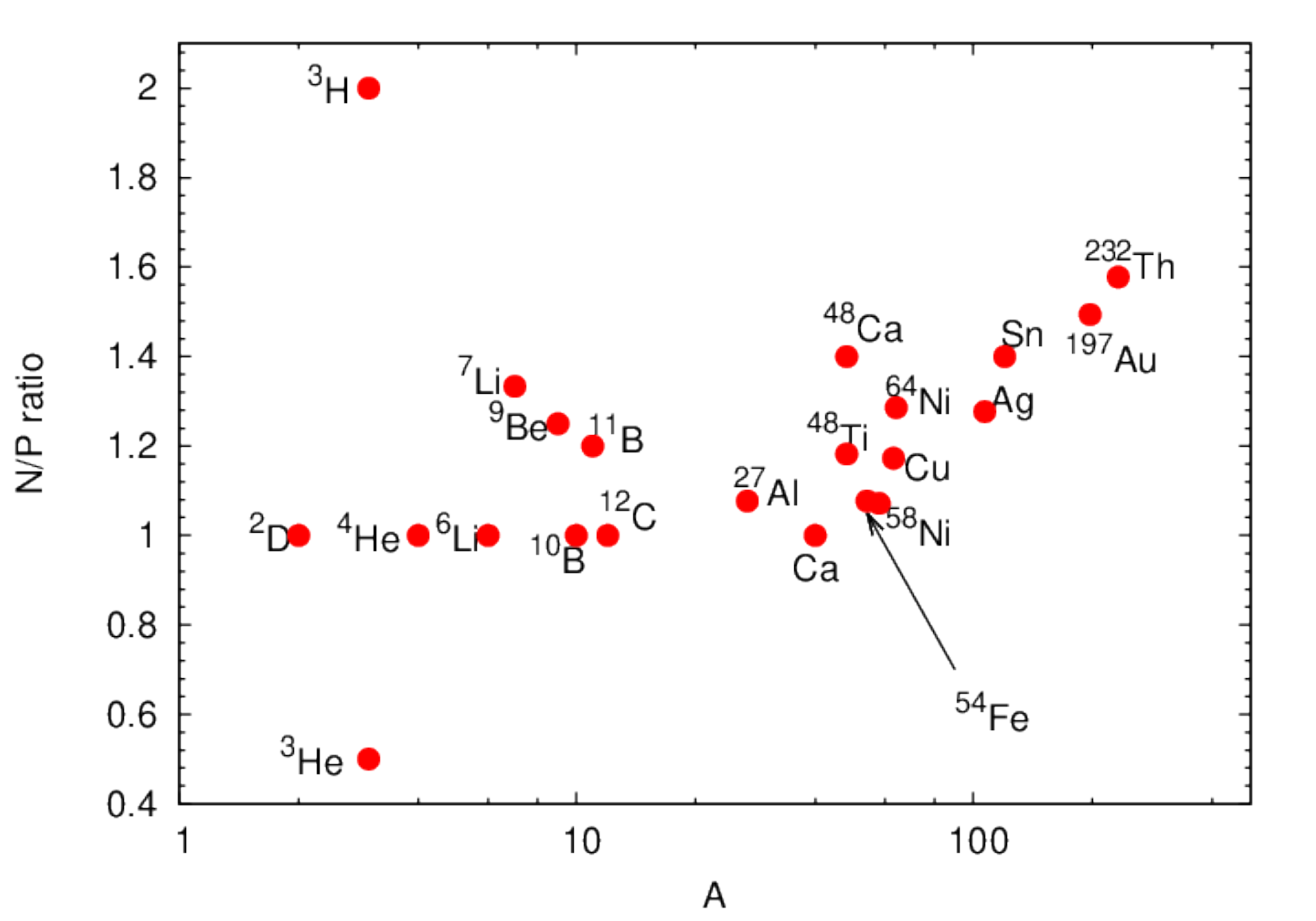}
    \caption{$N/P$ ratio vs. $A$ for nuclei that will be measured by JLab experiments E12-10-008 (EMC effect), E12-06-105 ($x>1$), and E12-10-103 (MARATHON) to further elucidate the apparent link between the EMC effect and short range  correlations.}
\label{fig:np_ratios}
\end{figure}

The above will be explored by JLab experiment E12-10-008~\cite{12gev_emc} in combination
with E12-06-105~\cite{12gev_xgt1}, E12-10-103~\cite{mar}, and E12-14-011~\cite{tritsrc}. E12-10-008 will measure the EMC effect for a wide range of nuclei, Fig.~\ref{fig:np_ratios},
aimed at elucidating the EMC-SRC connection, providing first measurements for a variety of light nuclei,
and exploring in-medium $N/P$ ratios via measurements of $A$ and $A\pm1$ nuclei.  E12-06-105 will make
measurements at $x>1$ in the two-nucleon correlation region for the same nuclei as well as make additional
measurements in search of possible three-nucleon correlations. Figure~\ref{fig:np_ratios} shows the $N/P$
ratio against $A$ for the nuclear targets that will be used for both experiments.
JLab Hall A has also recently completed a campaign of data taking on the mirror nuclei $^3$H and $^3$He. 
E12-10-103 will study the ratio of the structure functions between the two nuclei and 
E12-14-011 will study their short-ranged correlations.

\section{Isovector EMC Effects\label{sec:ivemc}}
One aspect of the EMC effect that has not been fully explored are isovector-dependent effects.  Such effects occur in nuclei with $N \neq Z$, e.g. many of those shown in Fig.~\ref{fig:np_ratios}, and would necessarily include degrees of freedom beyond the density, $\rho$, or nuclear mass number $A$, which has long been a method of parametrization~\cite{Malace:2014uea}.  In this situation, the nuclear $u_A$ and $d_A$ distributions can be modified separately, for example in an isovector mean field background or due to a preference in nucleon flavors in short range correlation pairing.  In a flavor separation which makes the assumption that $u_p = d_n$ and $d_p = u_n$ in the modified system, this manifests itself as an apparent charge symmetry breaking~\cite{Londergan:2009kj}, though only by the assumption that protons and neutrons retain their individual charge symmetry~\cite{Londergan:2009kj}. This assumption is not necessarily true for a bound system and can be invalid for asymmetric nuclei in the same vein as the binding energy has a symmetry energy component.  Continuing the binding energy analogy, as the symmetry energy is subleading to other bulk effects, such an isovector effect would be sub-leading to isoscalar modification effects.

The present world data have poor constraints on such an effect, in particular because many measurements of the EMC effect use symmetric or weakly asymmetric nuclei. One calculation~\cite{Cloet:2009qs,Cloet:2012td} using the Nambu-Jona-Lasinio (NJL) model and including the nuclear symmetry energy as an input, predicts deviations from the isoscalar EMC effect in the parton distribution functions on the order of several percent at large $x$.

Calculations based on SRCs predict a similar picture~\cite{Sargsian:2012sm, Arrington:2015wja}.  The observed correspondence of the short range correlation plateau with the slope of the EMC effect would indicate local densities as a driving mechanism~\cite{Weinstein:2010rt}.  Using the observation that proton-neutron pairs are found much more frequently than neutron-neutron or proton-proton pairs~\cite{Subedi:2008zz} and simple counting arguments, modification of protons and neutrons will be different in nuclei depending on the asymmetry.  In either a mean-field or SRC model, observation in the difference of quark flavors would require very high precision electromagnetic deep inelastic scattering measurements due to the suppression of the $d$ quark components weighted by the square of the electric charges. 

Weak interactions offer a novel method to probe flavor-dependent effects.  In charged-current processes, $u$ and $d$ quarks only participate in $W^-$ or $W^+$ exchange respectively. The NuTeV experiment~\cite{Zeller:2001hh} was carried out using neutrino beams at Fermilab and analyzed using the Paschos-Wolfenstein relation~\cite{Paschos:1972kj} to measure the weak mixing angle, $\sin^2\theta_\mathrm{W}$.  Due to the small neutrino cross section, heavy targets (iron) were employed which have a small neutron excess.  Charge symmetry in the bound nucleons was assumed and a significant deviation in $\sin^2\theta_\mathrm{W}$ was observed.  With the inclusion of an effect predicted by the NJL calculation noted above, much of the deviation is resolved~\cite{Cloet:2009qs,Bentz:2009yy}.  Additionally, this significantly augments neutrino scattering data which probe different flavor combinations~\cite{Schienbein:2009kk}.

%
The interference between electromagnetic and neutral currents through parity-violating DIS provides a complementary process to pure electromagnetic DIS, and when used together provide a powerful method to access the flavor dependence of the EMC effect~\cite{Cloet:2012td}. In parity-violating DIS, a polarized lepton beam scattered from an unpolarized asymmetric nuclear target will form a small parity-violating cross-section between the two beam helicity states, $\sigma_{L,R}$, which at leading order is 
\begin{equation}
\frac{ \sigma_R - \sigma_L }{\sigma_R + \sigma_L} = -\frac{G_F\,Q^2}{4 \sqrt{2}\, \pi\, \alpha} 
\left[ Y_1(y)\,a_1(x) + Y_3(y)\,a_3(x) \right],
\label{eq:phy:apv}
\end{equation}
where $G_F$ is the Fermi constant and $\alpha$ the electromagnetic fine structure constant.  The $Y$ functions are
\begin{equation}
Y_1(y) \approx 1, \qquad Y_3(y) \approx \frac{1 - (1-y)^2}{1 + (1-y)^2},
\end{equation}
and
\begin{eqnarray}
a_1(x) &= \frac{2\,\sum_q C_{1q}\, e_q\left[q(x) + \bar{q}(x)\right]}
{\sum_q\, e_q^2\left[q(x) + \bar{q}(x)\right]}, \\
a_3(x) &= \frac{2\,\sum_q C_{2q}\, e_q\left[q(x) - \bar{q}(x)\right]}
{\sum_q\, e_q^2\left[q(x) + \bar{q}(x)\right]},
\end{eqnarray}
where $y=\nu/E$, $E$ is the beam energy, $e_q$ is the quark electric charge couplings for flavor $q$, and $C_{1q}$ and $C_{2q}$ are the effective quark couplings dependent on the weak-mixing angle $\sin^2\theta_\mathrm{W}$~\cite{Patrignani:2016xqp}, with $C_{1u} \approx -0.19$ and $C_{1d} \approx 0.34$. The $a_1(x)$ term is dominant for fixed target, forward angle kinematics. 

\begin{figure}[tbp]
\centering\includegraphics[width=\columnwidth]{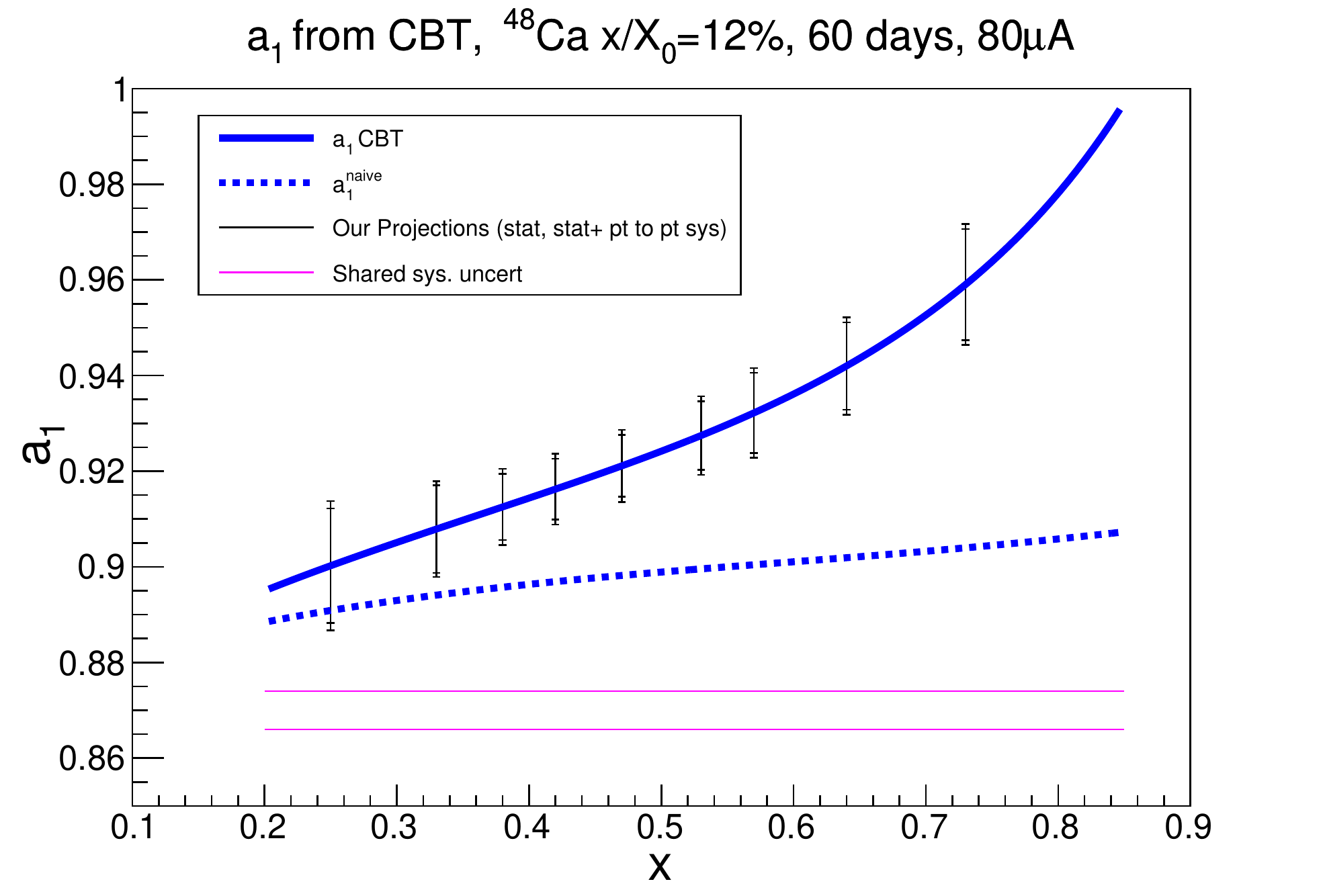}
\caption{Projected sensitivities of the quantity $a_1$ for a proposed parity-violating DIS experiment on a ${}^{48}$Ca target~\cite{emcpvdis}. The solid line is the full result from Ref.~\cite{Cloet:2012td} for $a_1$ in ${}^{48}$Ca, and the dashed line is the result when flavor-dependent nuclear effects are neglected.}
\label{fig:ivemc:pvdis}
\end{figure}

The first predictions for $a_1(x)$ for $N \neq Z$ nuclei was made in Ref.~\cite{Cloet:2012td}. These calculations, which included a self-consistent isovector mean-field whose strength was fixed by the symmetry energy, found that $a_1(x)$ is particularly sensitive to flavor dependent nuclear effects such as a flavor dependent modification of the nuclear parton distribution functions. The solid line in Fig.~\ref{fig:ivemc:pvdis} presents the result $a_1(x)$ from Ref.~\cite{Cloet:2012td} for ${}^{48}$Ca, and the dashed curve is the result with no flavor dependent nuclear effects. An experiment has been proposed~\cite{emcpvdis} for the SoLID spectrometer at Jefferson Lab would be able to test the prediction from Ref.~\cite{Cloet:2012td} to better than 5-$\sigma$ with a ${}^{48}$Ca target. The projected errors for this experiment are shown in Fig.~\ref{fig:ivemc:pvdis}.

\section{Polarized EMC Effect\label{sec:pemc}}
Since the discovery of the EMC effect~\cite{Aubert:1983xm,Bodek:1983ec} it has been clear that it embodies important new information about nuclear structure~\cite{Geesaman:1995yd}. However, this information is so well ingrained that there is as yet no consensus on what it is telling us. From one point of view it has been vigorously argued (see for example Refs.~\cite{Mineo:2003vc,Thomas:2016bxx,Guichon:2018uew}) that the model independent fact that there is a very strong attractive scalar mean-field in nuclei leads to changes in the internal structure of the bound nucleons, and that these changes not only explain the EMC effect but also play a vital role in the binding of atomic nuclei~\cite{Stone:2017oqt,Stone:2016qmi}. In order to distinguish between the various proposals that have been made by way of explanation for the EMC effect it is vital to find new observables which may shed light on which is correct.

One such proposal from Refs.~\cite{Cloet:2005rt,Cloet:2006bq} is to explore the spin-dependence of the EMC effect via measurement of the spin-dependent structure functions of atomic nuclei. This polarized EMC effect can be defined by
\begin{equation}
\Delta R_A(x) = \frac{g_{1A}(x)}{P_A^p\,g_{1p}(x) + P_A^n\,g_{1n}(x)},
\label{eq:pemc}
\end{equation}
where $g_{1A}(x)$ is a spin-dependent nuclear structure function, $g_{1p}$, $g_{1n}$ are the free nucleon structure functions and $P_A^p$, $P_A^n$ are the effective polarization of the protons and neutrons, respectively, in a nucleus of mass number $A$. There is an approved experiment at Jefferson Lab to measure the polarized EMC effect using $^7$Li~\cite{jlabspin}, and other nuclei include $^{11}$B, $^{15}$N, and $^{27}$Al. Another promising pathway at both Jefferson Lab and an EIC would be a detailed study of the complex of polarizable nuclei: $^1$H, $^2$H, $^3$H and $^3$He.

\begin{figure}[tbp]
\centering\includegraphics[width=\columnwidth]{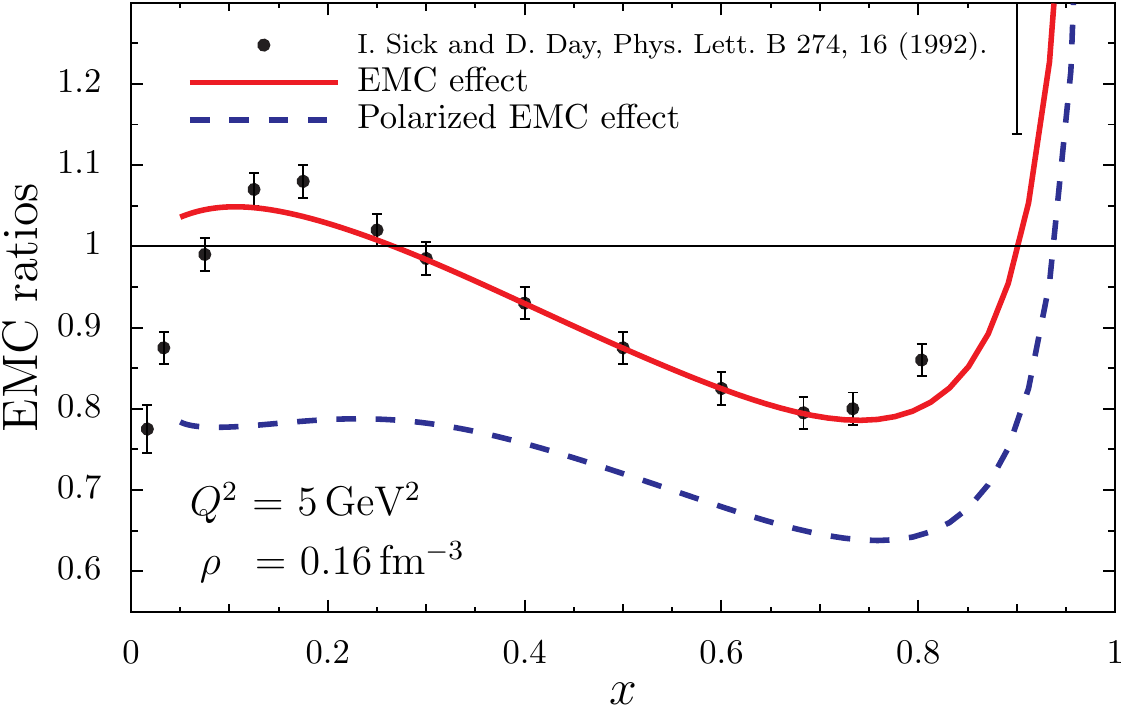}
\centering\includegraphics[width=\columnwidth]{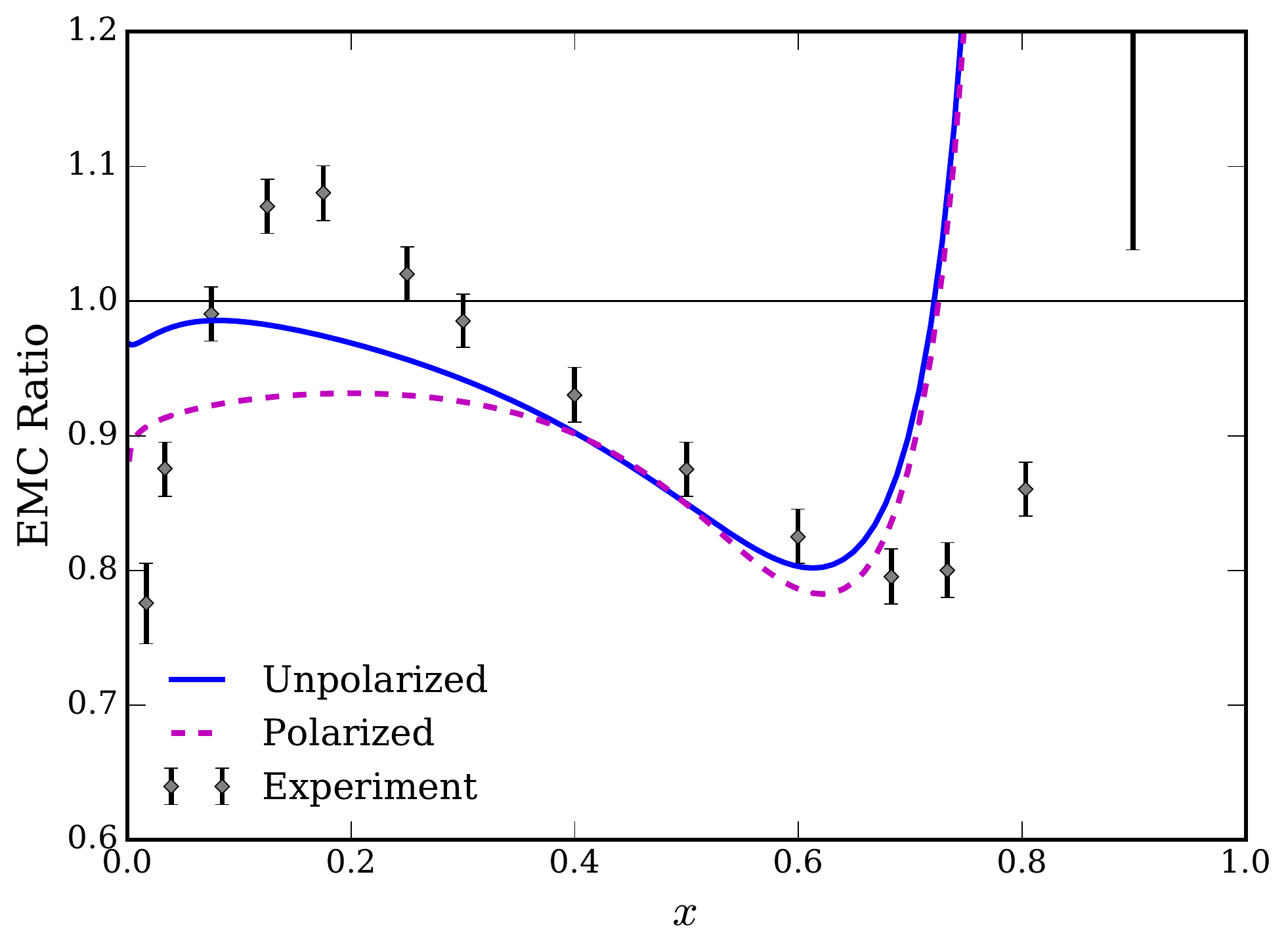}
\caption{{\it Top panel:} Unpolarized and polarized EMC effect results from Ref.~\cite{Cloet:2005rt}, obtained using a self-consistent NJL model calculation for isospin-symmetric nuclear matter. {\it Bottom Panel:} Analogous results from Ref.~\cite{Tronchin:2018mvu} obtained using the QMC model. In both cases the EMC data for nuclear matter is taken from Ref.~\cite{Sick:1992pw}.}
\label{fig:EMC_Com}
\end{figure}

The first calculations of the polarized EMC effect were for the hypothetical case of a polarized bound proton in nuclear matter~\cite{Cloet:2005rt}, where $g_{1A}(x) = g_{1p}^*(x)$ is the spin structure function of the bound proton which was assumed 100\% polarized ($P_A^p=1$ and $P_A^n = 0$). The effect of the medium was found to be dramatic~\cite{Cloet:2005rt} and is illustrated in the top panel of Fig.~\ref{fig:EMC_Com}, where the EMC effect was roughly twice as large for the spin-dependent case as for the unpolarized case.  For comparison, if one were to take the extreme position that all of the EMC effect arises from high momentum (highly correlated) nucleons~\cite{Weinstein:2010rt}, it seems unlikely that there would be any effect on the spin structure function.

The first calculation of the EMC effect within a self-consistent treatment of the modified structure of a bound nucleon was carried out 30 years ago using the quark meson coupling (QMC)  model~\cite{Thomas:1989vt}, however until very recently there had been no calculation of the polarized EMC effect within that model~\cite{Guichon:1995ue}. This is now of particular interest because it has been possible to derive an energy density functional equivalent to the QMC model, which starts with the modification of the quark structure of the nucleon in-medium, which has proven rather effective in nuclear structure studies~\cite{Guichon:2018uew,Stone:2017oqt,Stone:2016qmi}. 

Using the QMC model a calculation of the polarized EMC effect has now been completed by Tronchin {\it et al.}~\cite{Tronchin:2018mvu} and the result is reproduced in the bottom panel of Fig.~\ref{fig:EMC_Com}. We see that the prediction for the polarized EMC effect in the QMC model is about the same as that of the unpolarized effect, and a similar result was obtained in Ref.~\cite{Smith:2005ra} which used the chiral quark soliton model to perform an analogous calculation. The reason for the differences between both these calculations and the NJL result is currently not clear. However, a polarized EMC effect of the same size, or larger, than the unpolarized case would be difficult to explain using traditional nuclear structure effects or SRCs, because such an effect occurs in the valence nucleons. 

It has been suggested in Ref.~\cite{Thomas:2018kcx} that if SRCs are indeed the sole source of the EMC effect, there should be little or no polarized EMC effect. This renders the up-coming measurement of the structure function of a polarized $^7$Li target a key test of this fundamental issue in modern nuclear physics. As explained earlier, the dominant source of high momentum nucleons in nuclei (i.e., well above the Fermi level) is tensor correlations involving a neutron-proton pair in the $^3S_1$--$^3D_1$ state. The angular momentum barrier means that the highest momentum nucleons in a correlated pair will be in $D$-wave. That is, they may meet with low relative momentum in a shell model configuration, where their relative angular momentum is $S$-wave, but through SRCs they will be scattered into a high relative momentum $D$-wave state by the tensor force. Then one can show that the effective polarization of a valence nucleon once it has scattered through a SRC into a high momentum state will be of order $-10$ to $-15$\% instead of +100\%~\cite{Thomas:2018kcx}. Thus the structure function of this weakly polarized proton cannot exhibit a strong spin dependence. 

As SRCs will depolarize the valence proton, it is clear that were the EMC-type modification of structure functions to arise {\em only} through the change of the structure function of a high momentum, far off-mass-shell nucleon in a correlated pair, there can be very little EMC effect on the nuclear spin structure functions. Once one divides the measured nuclear structure function by the effective nucleon polarization one expects little nuclear modification of the spin structure function and hence no significant spin-dependent EMC effect in this model. The polarized EMC effect is then a clear example of a major difference in the predictions of the mean-field and SRC explanations of the EMC effect.

We stress that for the polarized EMC effect results shown in Fig.~\ref{fig:EMC_Com} the proton is defined to be 100\% polarized and is embedded in nuclear matter. For a real nucleus, such as $^7$Li for which a measurement is planned at Jefferson Lab~\cite{jlabspin}, one needs to account for the reduced local density of the valence nucleons and the fact that the polarization of the bound proton is less than that of the nucleus. For the case of $^7$Li nuclear structure calculations, including variational Monte Carlo, find $P_A^p = 0.87$ and $P_A^n = 0.04$~\cite{Pudliner:1997ck}.

\section{Drell-Yan and Nuclear PDFs\label{sec:DY}}
Most of our knowledge on the medium-modified PDFs come from DIS experiments. Although extensive and precise, these data are not able to independently investigate the nuclear effects on sea and valence quarks. Moreover, they do not discriminate between the different flavors participating in the reaction. Finally, they do not probe the nuclear gluon distributions. Additional studies of the effect of the medium on the PDFs can be provided by dedicated fixed-target dimuon production experiments. In the Drell-Yan mass region  these experiments can be used to isolate valence and sea distributions as well as to separate the two light flavors of the quark PDFs~\cite{Bickerstaff:1985ax}. The unknown gluon distributions in nuclei can be accessed by analyzing the contributions to the charmonium production cross-sections. 
 
\subsection {Valence quark distributions and the EMC effect}
\begin{figure}[tbp]
  \centering\includegraphics[width=\columnwidth]{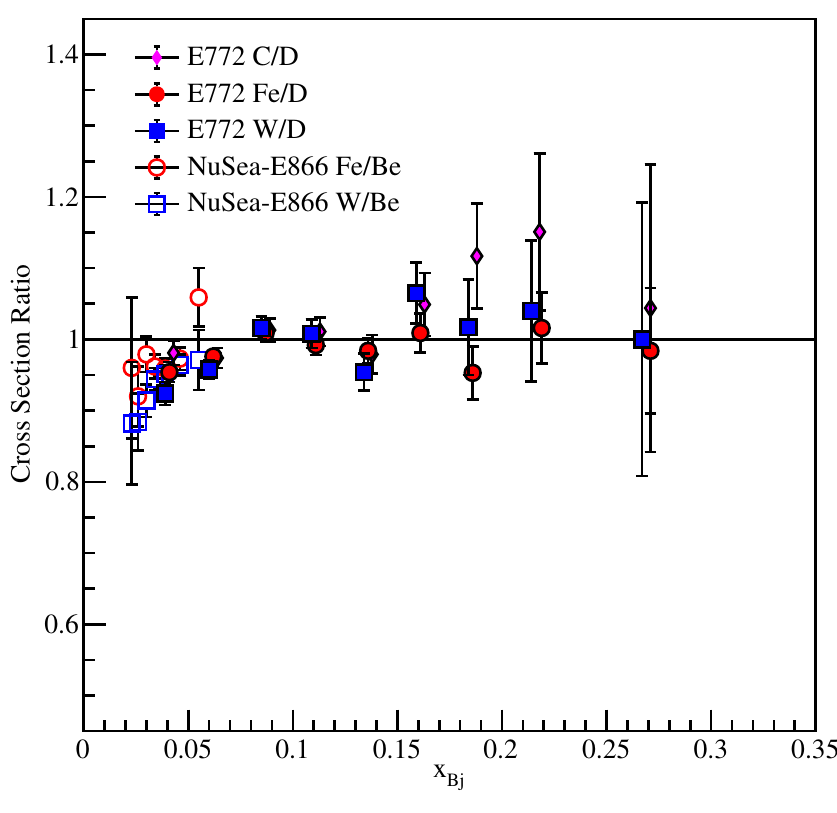}
  \caption{Drell-Yan cross section ratios from E772 and E866 \cite{Alde:1990im, Vasilev:1999fa}.}
  \label{fig:drell-yan}
\end{figure}

Only a few Drell-Yan experiments have investigated the medium modifications in nuclear targets. 
The Drell-Yan data from the E772 experiment~\cite{Alde:1990im} show no visible nuclear effects in 
the anti-shadowing domain, both E772 and E866 experiments~\cite{Vasilev:1999fa} are compatible 
with the DIS data in the onset of the shadowing region, Fig.~\ref{fig:drell-yan}.  An exploration of the medium effects for larger $x$ values was recently completed by the SeaQuest experiment~\cite{seaquest} at Fermilab. 

All these experiments use proton beams and they are mainly sensitive to the $\bar{u}$
distributions in nuclei.  Since pions contain also valence antiquarks, pion beams are  
an ideal tool for the investigation of the valence quark distributions. Pion-induced Drell-Yan data have 
been taken in the past by two CERN experiments, NA3~\cite{Badier:1981ci} 
and NA10~\cite{Bordalo:1987cs}.  For both experiments the achieved statistics is limited
and the conclusions of the authors inconsistent.  

A dedicated Drell-Yan experiment, 
combining nuclear and deuterium targets could significantly improve the results. 
A high-intensity pion beam with incident momenta between 100 and 200~GeV/$c$ is presently available 
only at CERN. Since in Drell-Yan experiments the total cross section increases with the energy, a higher 
energy is preferable. On the other hand, a maximal coverage of the high-$x$ EMC region could require a 
lower incident momentum.  

\subsection{Flavor dependence of the EMC effect} 

The sensitivity of the available Drell-Yan data on the flavor-dependent effect was explored in 
Ref.~\cite{Dutta:2010pg}. A comparison with calculations based on the model of Ref.~\cite{Cloet:2009qs} 
shows that the  statistics of the available data~\cite{Badier:1981ci,Bordalo:1987cs} is insufficient to draw firm conclusions. 
The flavor dependence may also have an effect 
on the global nuclear PDF fits presently available. Indeed, when releasing the flavor constraints in these fits, 
the model uncertainties increase~\cite{Paakkinen:2016wxk} by a large factor. 

The effect coming from the flavor dependence can be further amplified by 
comparing positive and negative pion beam Drell-Yan data on the same heavy target. Negative pions probe mainly the 
target $u$-valence quarks, whereas positive pions prefer $d$-valence quarks. Pion beams of both charges
are presently available only at CERN. 

\subsection{Nuclear dependence of the gluon distribution}
The effect of the nuclear mean-field on the gluon distribution is presently unknown. Available nuclear PDFs produce very different results~\cite{Cazaroto:2008qh} in the whole region of $x$ and particularly for $x > 0.1$. The J/$\psi$ production data,  usually collected in parallel with Drell-Yan data, can be used as a probe of the nuclear gluon distributions.  The production of J/$\psi$ proceeds through either the $q\bar q$ or the $gg$ fusion processes, the $gg$  contribution being the dominant one~\cite{Vogt:1999dw} for $x_F = x_2 - x_1 < 0.5$  even at low center-of-mass energies. 

As for Drell-Yan, most of the J/$\psi$ production data were taken with proton beams. The nuclear dependence  was primarily used for the evaluation of the absorption of the J/$\psi$ as a function of the atomic number.  Pion-induced J/$\psi$ data have been taken by several experiments, but most of the time with a single target.  A comparison of several nuclear targets (Si, Cu, W) was also made~\cite{Alexandrov:1999ch},   but unfortunately for the total cross-sections only. There is some data for high-statistics J/$\psi$ production data on the $x$ dependence, combining several nuclear and light targets. 

The strong-interaction J/$\psi$ production process has one important advantage: its cross-section is large.  Large statistics J/$\psi$ production data can be collected and  used to precisely map out the $x$-dependence of the nuclear modifications. Such data could be  used for a better assessment of the charmonium production process.  The insufficient knowledge of this process is presently the main source of systematic uncertainties, preventing reliable extraction of the nuclear gluon distributions.  The gluon distributions could  therefore be inferred with minimized model-dependence uncertainties. Interestingly, J/$\psi$ production can also be used to study charge symmetry violation~\cite{Piller:1995nc}.

\section{Opportunities with low $\mathit{A}$ nuclei\label{sec:light}}
The difficulties in describing nuclei within QCD are myriad, e.g., all the challenges of understanding nucleon structure from QCD are inherited for nuclei, and usually amplified, and effective field theory approaches must deal with few- and many-body systems with numerous new energy scales. While QCD inspired models for nuclear structure can be applied across the table of nuclides~\cite{Cloet:2015tha,Stone:2017oqt,Stone:2016qmi,Cloet:2006bq,Modarres:2018ymh}, lattice QCD techniques are currently only able to approach descriptions of nuclei with $A \leq 4$~\cite{Chang:2015qxa}, which is at the boundary of where one may expect large nuclear modification effects to occur. To move toward a more complete QCD description of nuclei advancement in QCD inspired models and {\it ab initio} techniques is critical, together with a robust comparison to few-body nuclear data. A key example going forward is provided by the formalism of generalized parton distributions, which can be extended to nuclei, and offer a new window of exploration which unifies the study of the medium modification of PDFs and form factors, and provides the opportunity to obtain a quark-gluon tomography of nuclei. 

The deuteron is the lightest non-trivial nucleus and it has a wave function dominated by a loosely bound proton--neutron configuration.  As such, it offers a range of possibilities to study QCD phenomena at hadronic and partonic scales.  As it offers a (bound) neutron target, DIS off a deuteron is together with parity-violating DIS the main tool to perform flavor separation of quark distribution functions.  As a bound $pn$ system, it also offers a window into the QCD origin of the nucleon-nucleon force and medium modifications of nucleon properties~\cite{Boeglin:2015cha}.  In kinematics where a high-energy probe interacts with both nucleons in the deuteron, coherent phenomena in QCD such as shadowing and saturation can be studied.  Lastly, the deuteron is a spin-1 hadron and as a result studies with a polarized deuteron offer additional observables and opportunities beyond those of the free polarized nucleon.

When inclusive DIS is carried out on a nucleus, the process averages over all initial configurations.  This means one has to account for possible medium modifications and include binding effects and non-nucleonic degrees of freedom in the nuclear wave function.  A handle on the control of the initial state of the target nucleus is provided by the spectator tagging process, where a slow nucleon (relative to the center of mass of the nucleus) is detected in the target fragmentation region of the final state.  For the deuteron, controlling the initial target state through spectator tagging has several advantages.  Spectator tagging effectively identifies the active nucleon participating in the DIS reaction (proton tagging enables neutron structure studies and the other way around) and suppresses nuclear binding effects at low spectator momenta. 

\begin{figure}[tbp]
\centering\includegraphics[width=\columnwidth]{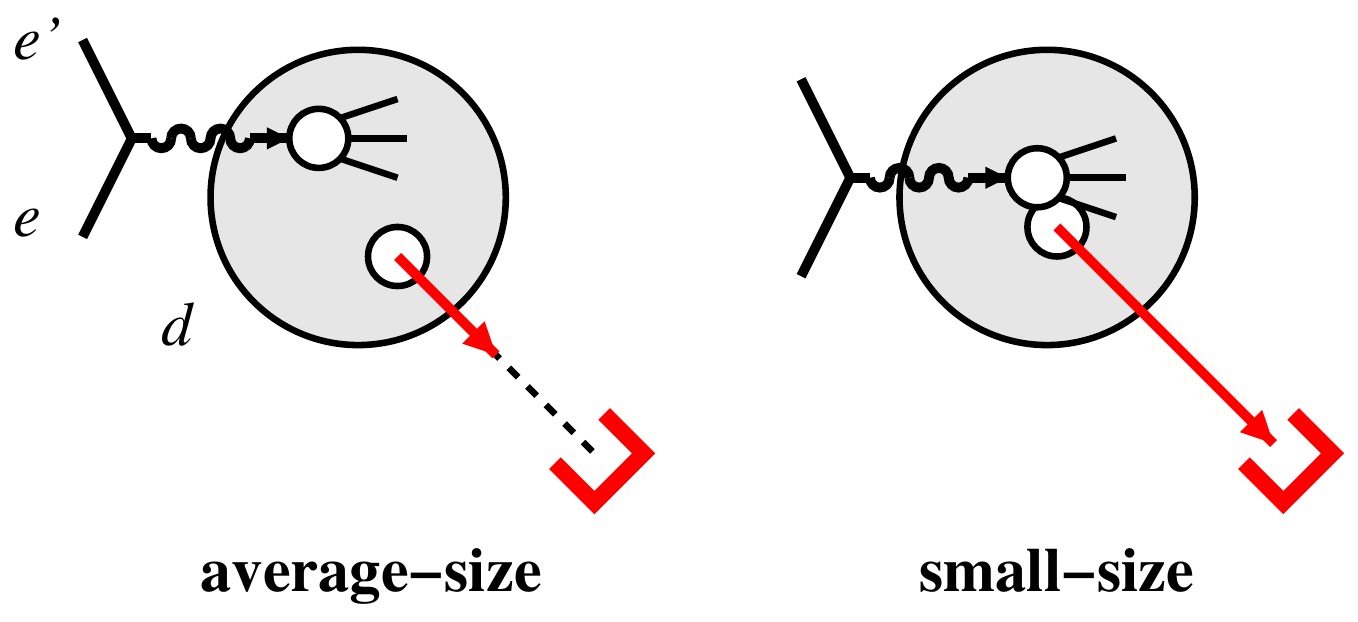}
\caption{Deuteron lab frame depiction of the spectator tagging process.  Low spectator momenta select average-sized configurations, large spectator momenta access small-sized ones~\cite{deutLDRD}}
\label{fig:size}
\end{figure}

By varying the momentum of the detected spectator, one can also select compact (high momentum) or loose configurations (low momentum) of the deuteron, Fig.~\ref{fig:size}.  This allows the study of nuclear binding effects and the nature of the $NN$-interaction at different length scales and densities.  Of course, one has to account for the possible final-state interactions (FSIs) between the spectator and the DIS products.  The deuteron has the advantage there is only one possible spectator nucleon, meaning FSIs are tractable.  Moreover, in kinematics with large FSIs they can be used to obtain information about the process of hadronization in the DIS products~\cite{Cosyn:2017ekf}.  

The deuteron also has the advantage that first principles non-relativistic wave functions are available to perform theoretical calculations.  In high-energy reactions, the structure of the light-front deuteron wave function is known~\cite{Frankfurt:1981mk,Keister:1991sb} and it can be matched to the non-relativistic wave functions at small relative momenta.

To access on-shell nucleon structure, the technique of pole extrapolation can be applied to the spectator tagging process.  Here, observables as a function of spectator momentum are extrapolated in the unphysical region to the on-shell point of the active nucleon.  The small binding of the deuteron implies the extrapolation length into the unphysical region is quite small, resulting in controllable errors on the extrapolated values.  The no-loop theorem implies that although higher order diagrams (such as FSIs) contribute to the spectator tagging process, they do not contribute at the on-shell point~\cite{Sargsian:2005rm} implying the access to free neutron structure functions for proton tagging.  Additionally, at the on-shell point the deuteron polarization is almost 100\% transferred to both nucleons due to the deuteron $S$-wave dominance, enabling the extraction of on-shell neutron spin structure in spectator tagging with polarized beams.  

\begin{figure}[tbp]
\centering\includegraphics[width=0.6\columnwidth]{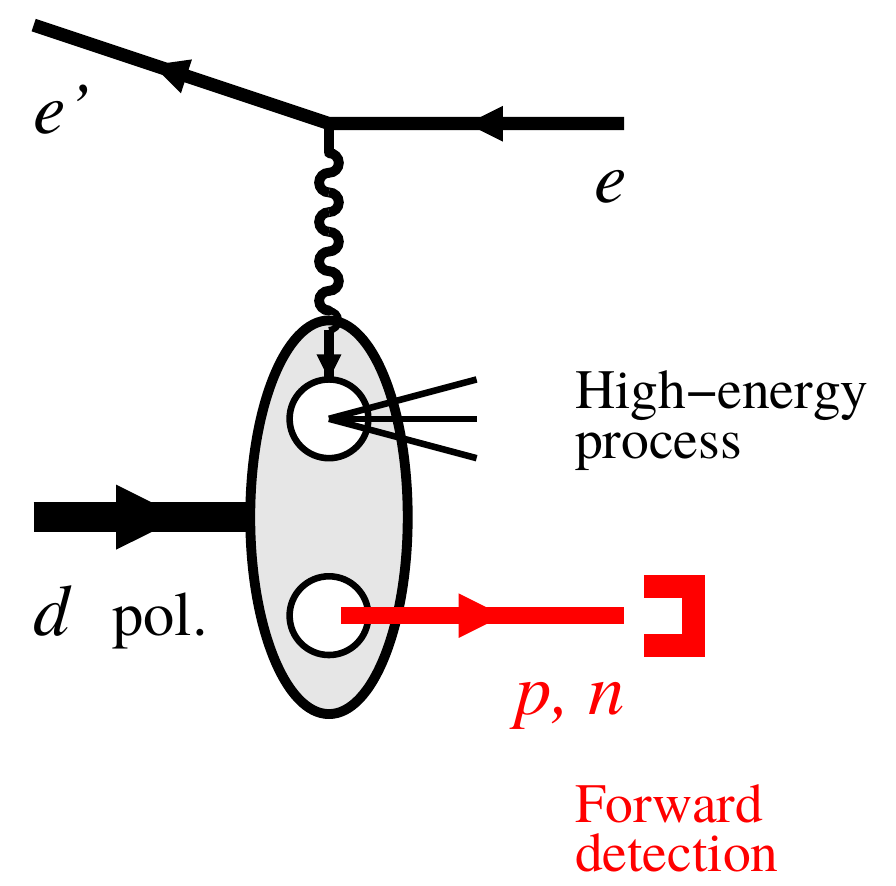}
\caption{Collider frame depiction of the spectator tagging process with forward detectors capturing the spectator nucleon~\cite{deutLDRD}}
\label{fig:collider}
\end{figure}

Spectator tagging on the deuteron has been measured at JLab at both high (DEEPS experiment~\cite{Klimenko:2005zz}) and low spectator momenta (BoNuS experiment~\cite{Baillie:2011za}). In a fixed target setup this measurement is challenging, especially at low spectator momenta. Hence a dedicated detector had to be installed in the BoNuS experiment.  These difficulties disappear for an electron-ion collider, where the spectators still move forward after the DIS event with momenta of the order of half the deuteron beam momentum, and can be detected in forward detectors, Fig.~\ref{fig:collider}.  Together with the wide kinematic range in Bjorken $x$ and $Q^2$ that an electron-ion collider offers, spectator tagging can make a significant impact on on-shell neutron structure data and studies of medium modifications.  Extensions of the tagging process are also possible using light nuclei beams beyond the deuteron and measuring more exclusive channels.  The theoretical framework and simulation tools for the spectator tagging process are under active development~\cite{deutLDRD,Guzey:2014jva,Cosyn:2016oiq} and estimates for the size of FSIs at an electron-ion collider have recently been provided~\cite{Strikman:2017koc}.

A straightforward way of probing the additional spin degrees of freedom the deuteron offers is by performing inclusive DIS off a tensor-polarized target, which is sensitive to four additional structure functions~\cite{Hoodbhoy:1988am}.  One of these $b_1$ has an interpretation in the parton model as a linear combination of unpolarized quark PDFs in a polarized spin-1 hadron
\begin{equation}
b_1=\frac{1}{2}\sum_q e_q^2\lf(q^0-q^1\rg),
\end{equation}
where $q^i$ is the quark PDF in a hadron with polarization $i$ along the virtual photon momentum.  Hence, $b_1$ probes the interplay of nuclear and quark degrees of freedom.  Currently, there exists only one measurement of $b_1$ from HERMES~\cite{Airapetian:2005cb} that cannot be explained by conventional deuteron convolution models.  A new measurement is planned at JLab~\cite{Slifer:2013vma}, which spurred updated convolution model calculations~\cite{Cosyn:2017fbo} and estimates of hidden color and pion cloud contributions~\cite{Miller:2013hla}.

\section{Poincar\'e Covariant Light-Front Spectral Function and Nuclear Structure\label{sec:lf}}
{The Poincar\'e covariant spin-dependent spectral function, proposed in \cite{PhysRevC.95.014001,Pace:2013bq,Scopetta:2014yoa,Pace:2016eiq} and based on  the light-front (LF) Hamiltonian dynamics \cite{Keister:1991sb,Dirac:1949cp},  is a useful tool for a correct relativistic treatment  of nuclear structure, suitable for the study of deep inelastic scattering (DIS) or semi-inclusive deep inelastic scattering (SIDIS) processes at high momentum transfer \cite{mar,sid1,sid2}.}
 {Indeed the Bakamjian-Thomas construction \cite{Bakamjian:1953kh} of the Poincar\'e generators
  allows one to embed the successful phenomenology for few-nucleon systems in a
Poincar\'e covariant framework.}

The LF spectral function
for a three-fermion system (as the $^3$He or a nucleon in valence approximation) depends on the  energy $\epsilon$ of the spectator subsystem  and on the LF  momentum $ \tilde{\bm \kappa}$ of the knocked out particle in the {{intrinsic reference frame of the (particle - spectator pair) cluster}}. It is built up from the overlaps of the ground eigenstate of a proper mass operator for the system \cite{PhysRevC.95.014001,Pace:2013bq,Scopetta:2014yoa,Pace:2016eiq}  and
  the tensor product of a plane wave for the particle
 times the fully interacting  state for the spectator.
 {{The use of the momentum {{${\tilde{\bm \kappa}}$}}
   allows one  to take care of macrocausality \cite{Keister:1991sb} and  to introduce
   {{a new effect of binding in the spectral function.}}}}

The LF
 spectral function
fulfills {normalization and momentum sum rule  at the same time. Integration of the spectral function on the energy
$\epsilon$
of the pair yields the LF spin-dependent momentum distribution that
can  be expressed through six scalar functions, straightforwardly obtained from the system LF wave function as integrals on the relative momentum of the spectator pair.

{{Calculations of DIS
 processes  based on a LF spectral function could indicate which is the gap with respect to the experimental data to be filled by effects of non-nucleonic degrees of freedom or by modifications of nucleon structure in nuclei.}}

\section{Probing the nucleus with tagged reactions\label{sec:tagged}} 

\begin{figure}[tbp]
\centering\includegraphics[width=\columnwidth]{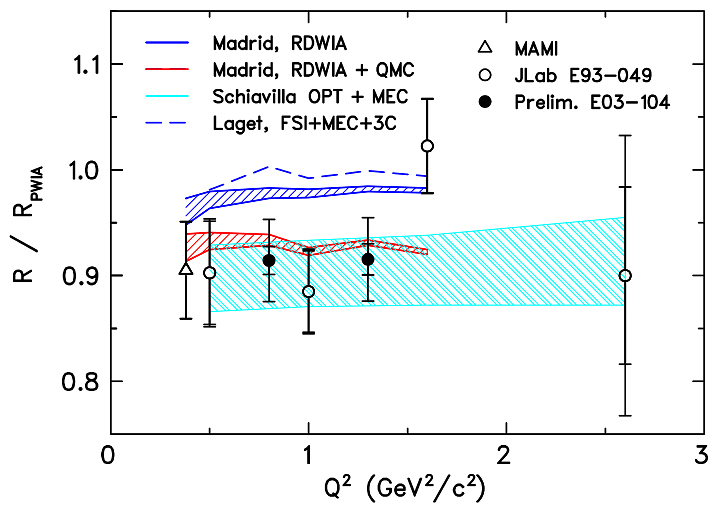} \\
\centering\includegraphics[width=\columnwidth]{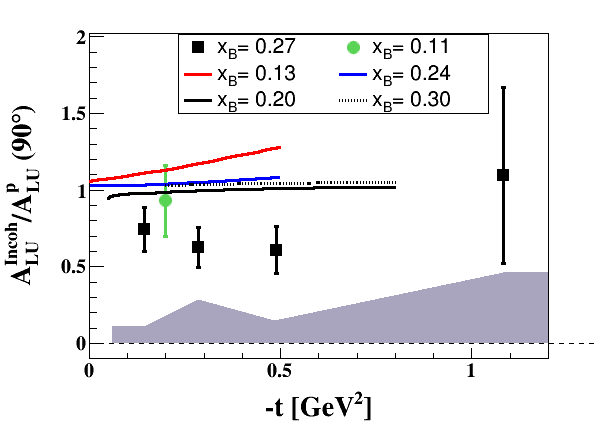}
\caption{Top: results of \cite{Strauch:2002wu} for the quasi-elastic form factor relative to
the plane wave impulse approximation. Bottom: CLAS preliminary results 
for the beam spin asymmetry in incoherent nuclear DVCS relative to the free proton DVCS.}
\label{fig:QEincoh}
\end{figure}

Medium modification is not limited only to differences in the PDFs,
but could also be manifest in other observables, such as the elastic nucleon form factors.
This idea
triggered several experiments to measure the quasi-elastic 
process ($e+A \rightarrow e+p+X$). In this process, the 
elastic scattering occurs on a bound nucleon and allows the extraction of its modified 
form factor. One of the expectations from these measurements was to detect a change
in size of the bound nucleon compared to the free one. This has also motivated 
deeply-virtual Compton scattering (DVCS) experiments, where a similar process is accessible, the so-called
incoherent nuclear DVCS ($e+A \rightarrow e+p+\gamma+X$). Results for these two 
channels are presented in Fig.~\ref{fig:QEincoh}, and show in both cases a 
significant deviation between the bound and the free nucleons. 
The difficulty with the interpretation of these measurements lies into the 
effect of final-state interactions. Indeed, the reaction products are likely
to re-interact with the remnants of the nucleus, and this affects significantly the
results. The calculation of these final state effect is complex and leads to large model 
uncertainties. 
Another problem is that in the calculation of these processes, it is important
that the initial and final-state nucleons are the same. This cannot be guaranteed
in a nucleus where one can have a off-shell nucleon in the initial state or 
have processes where a charge is exchanged and a neutron becomes a proton. For
these reasons, the debate remains open around 
the interpretation of the data on bound nucleon scatterings~\cite{Benhar:2006wy}.

The solution to these problems is the tagging method~\cite{CiofidegliAtti:1999kp}, in which the nuclear fragments
are detected as illustrated in Fig.~\ref{fig:size} for the simplest case, deuterium.
In this process ($e+d \rightarrow e+p_s+X$), the high-energy electron is measured
together with the low-energy proton. The measurement of the proton in the backward 
direction ensures that it was not part of the hard interaction, thus noted with 
the $s$ subscript for spectator, and transforms the deuterium into 
an effective neutron target. First results of such a measurement
have been reported in~\cite{Baillie:2011za} with the goal to extract the structure
function of the neutron. We show in Fig.~\ref{fig:wstar} a result of this experiment
comparing the invariant mass obtained with and without the tagging method. It is 
clear that the tagging method gives a much better resolution of the structure 
present in the invariant mass distribution. This successful measurement of a tagged process opens the way for more
experiments of the same kind in the future. 

\begin{figure}[tbp]
\centering\includegraphics[width=\columnwidth]{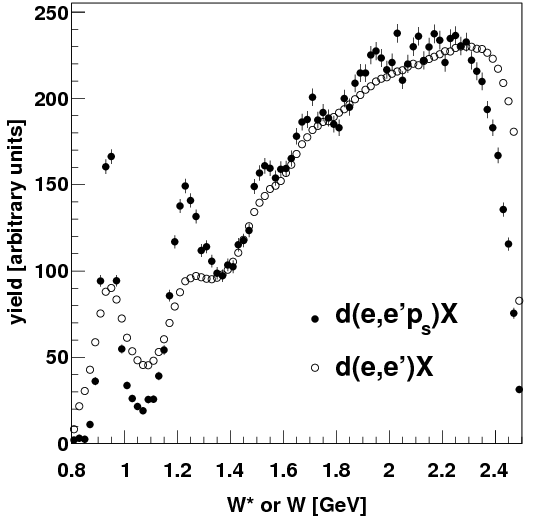}
\caption{Neutron-electron invariant mass obtained from tagging (full points) compared 
to the invariant mass obtained
from deuterium data (hollow points)~\cite{Baillie:2011za}.}
\label{fig:wstar}
\end{figure}

The nuclear remnant tagging is the extension of the deuterium tagging to heavier nuclei, which
consists in measuring the reaction ($e+A \rightarrow e+(A-1)_s+X$), with
$(A-1)_s$ the spectator remnant of the nuclear target. This measurement is very 
interesting as it gives a direct information on which nucleon was hit by the 
deep inelastic electron scattering in the nucleus. Also, by selecting 
low momentum backward emission of the $(A-1)_s$, we can suppress the 
final state interactions that are often a problem in such nuclear reactions. 
Detailed studies~\cite{CiofidegliAtti:2003pb,Alvioli:2006jd} have indeed shown 
that the final state interactions effects are minimized when the nuclear recoil
is detected in a backward angle relative to 
the virtual photon direction and maximized in perpendicular kinematics.
The detection of such recoil nuclei is however extremely challenging.

Tagging recoiling remnants has another interest in the quest to understand nuclear structure.
The kinematics of the nuclear remnants contain information on the
initial state of the nucleons in the nucleus. By performing at the same time the 
tagging and a deep inelastic scattering, we probe simultaneously the nucleon and the 
quark structure of the nucleus. Tagging is therefore a unique tool to relate the EMC
effect to more classical nuclear effects and observe if there is any correlation between
them. The more natural variable to use for these studies is the nucleon virtuality.
It can be calculated~\cite{CiofidegliAtti:2007ork} in the impulse 
approximation, where the nucleon momentum is exactly $\mathbf{p} = -\mathbf{P}_{A-1}$, 
giving
\begin{eqnarray}
    v(|\mathbf{p}|, E) &  =  & \left (M_A - \sqrt{(M_A - m_N + E)^2 + \mathbf{p}^2} \right )^2  \\* \nonumber
                    & & -  \mathbf{p}^2 - m_N^2,
\end{eqnarray} 
where $E$ is the removal energy, $M_A$ the mass of the target nucleus
and $m_N$ the mass of the nucleon. The nucleon virtuality is a key 
observable to understand the nuclear quark and gluon structure,
as there are radically different predictions for its impact on the partonic structure.
Indeed, the descriptions of the EMC effect based on nucleon dynamics
predict a strong correlation between virtuality and nucleon modification,
while the ones involving other hadronic degrees of 
freedom or mean field effects do not. 

Nuclear tagging measurements have never been performed in the past
on nuclei with $A>2$ due to detector limitations. The radial time projection chamber used 
for the experiments described above~\cite{Baillie:2011za}
is unable to differentiate isotopes and thus ensure the identification of 
nuclear remnants. There are plans by the CLAS collaboration to remediate this issue
explore new ground with the construction of the new ALERT detector to perform nuclear tagging measurements 
for the first time~\cite{Armstrong:2017zqr,Armstrong:2017zcm}.

\section{GPDs of Nuclei\label{sec:GPDs}}

It has been clear for many years that inclusive measurements alone cannot
provide a 
full quantitative explanation of the EMC
effect \cite{Aubert:1983xm}.
Imaging of nuclei, now possible for the first time
through DVCS and deeply-virtual meson production (DVMP), can answer the question using GPDs as a tool \cite{Dupre:2015jha}.
By comparing transverse spatial quark and gluon distributions in nuclei
or bound nucleons (to be obtained in coherent or incoherent DVCS, respectively) to
the corresponding quantities in free nucleons,
one can realize a pictorial representation of the EMC effect.
The presence of non-nucleonic degrees of freedom, 
as addressed in Ref. \cite{Berger:2001zb},
or the 
change of confinement radii in bound nucleons, will be observed. 

\begin{figure}[tbp]
\centering\includegraphics[scale=0.28]{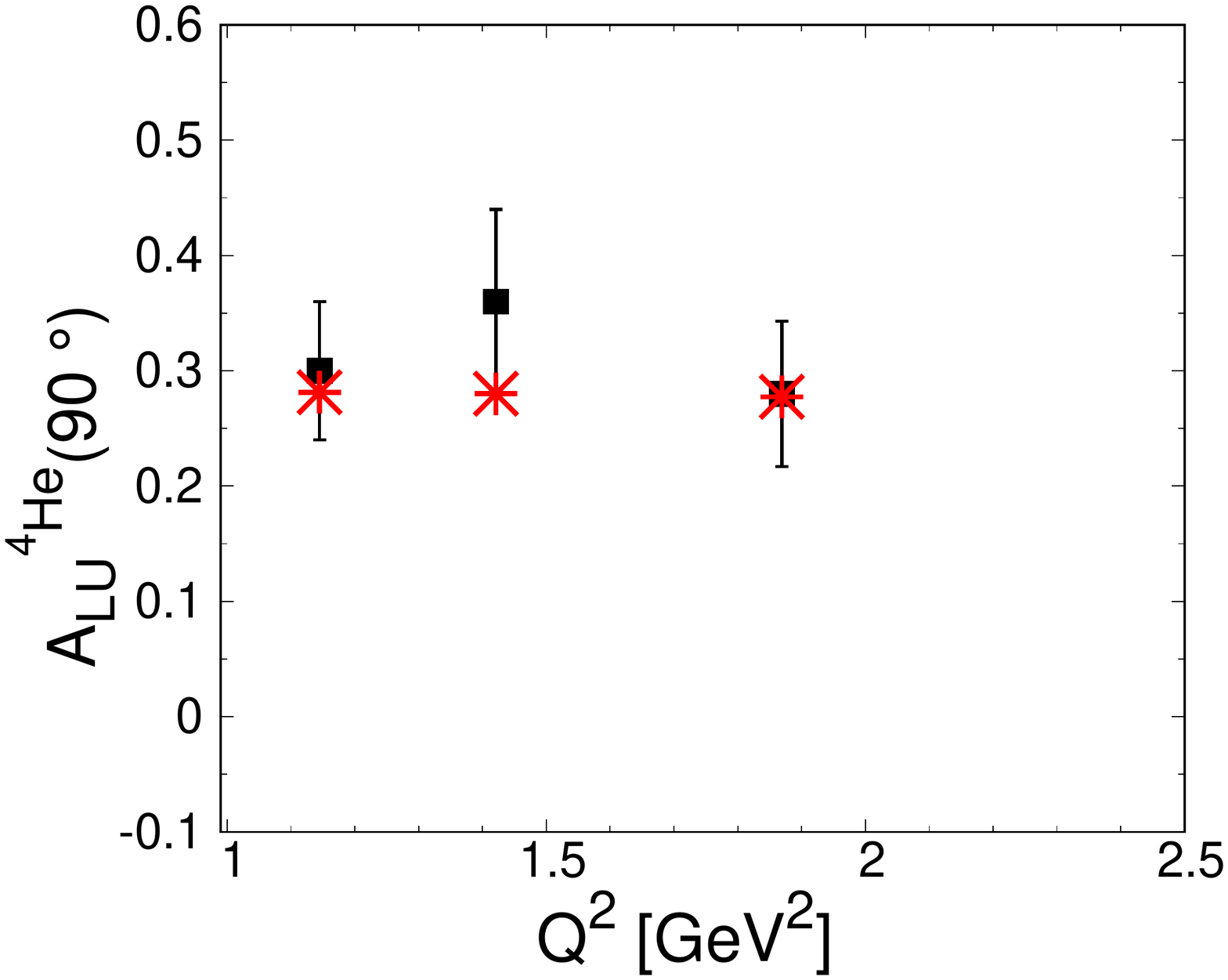}
\vskip -1.cm
\centering\includegraphics[scale=0.28]{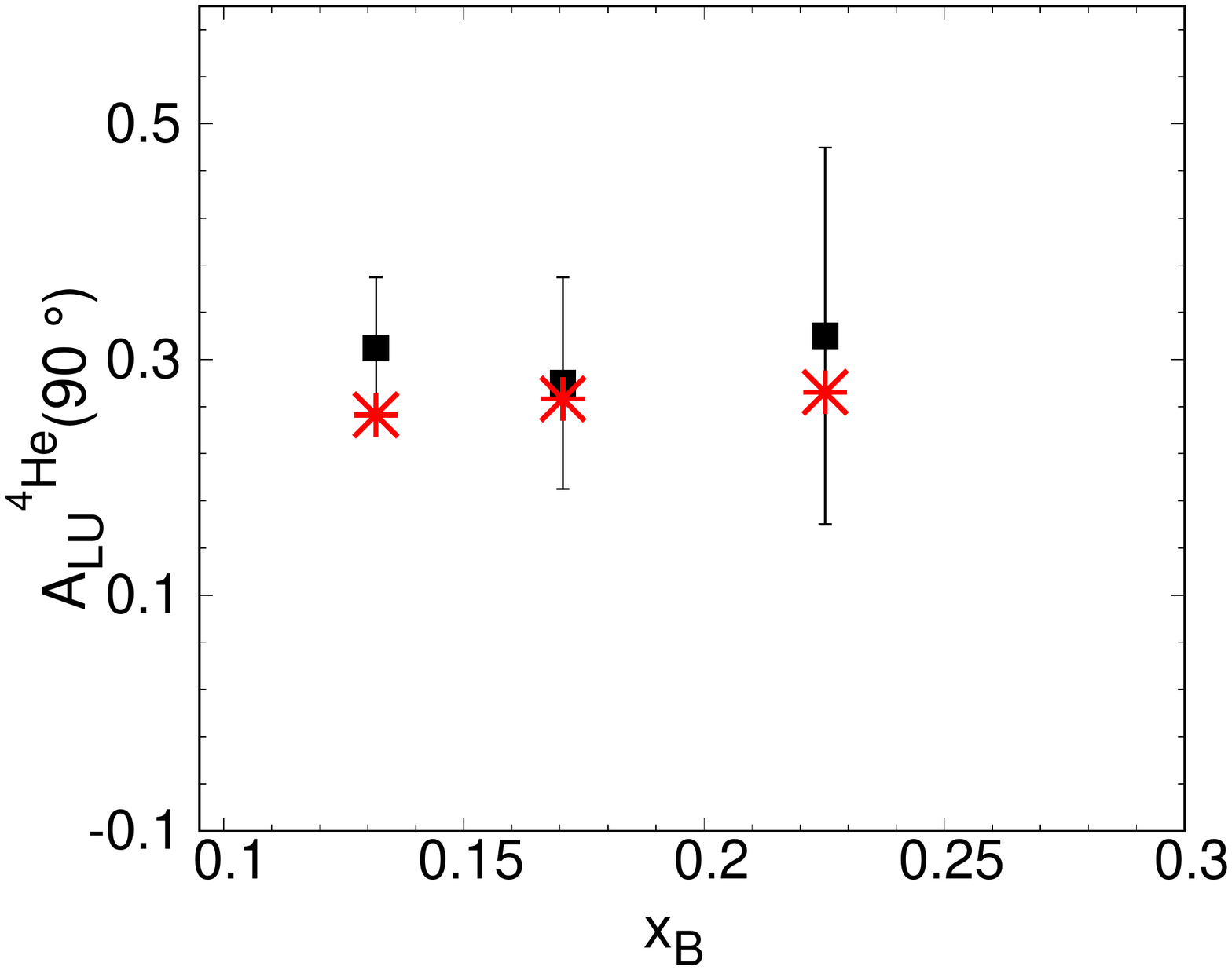}
\vskip -1.cm
\centering\includegraphics[scale=0.28]{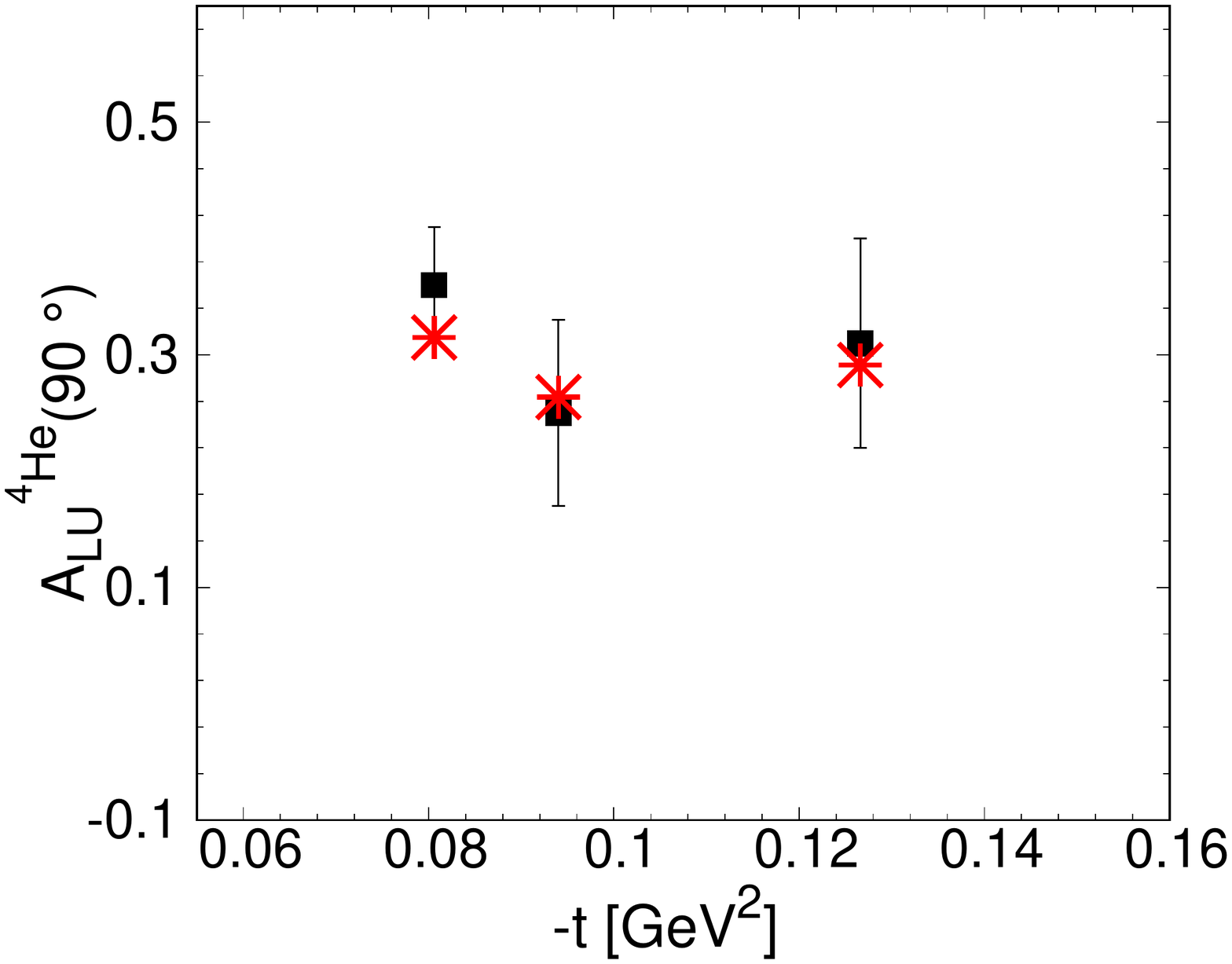}
\vskip -.5cm
\caption{(Color online) $^4$He azimuthal beam-spin asymmetry $A_{LU}(\phi)$,
for $\phi = 90^o$: results of Ref. \cite{PhysRevC.98.015203} (red stars) compared with data
(black squares)
\cite{Hattawy:2017woc}.
From top to bottom, the quantity is shown in the experimental
$Q^2$, $x_B$ and $t$ bins, respectively. 
}
\label{alu}
\end{figure}

The valence region at intermediate $Q^2$, investigated by 
JLab at 12~GeV, is crucial to access
the most discussed region of the EMC effect.
For light nuclei, such as $^2$H, $^3$He, $^4$He,
a sophisticated evaluation of conventional nuclear effects,
although sometimes challenging, is possible.
This would allow one to distinguish them
from exotic effects, which could be responsible for
the observed EMC behavior.
Without realistic benchmark calculations,  
interpreting experimental data cannot be conclusive.


Deuterium is very interesting, both for its rich spin structure
and for the possibility of extracting neutron information.
As a spin-one object, the deuteron admits a large number of GPDs,
as well as electromagnetic and gravitational form factors (GFFs),
which encode more information than is available in the proton or neutron alone.
For instance, the deuteron appears to have two $D$-terms which encode
the distribution of shear and pressure, while the nucleon has one.
Understanding how forces are distributed in the deuteron
will give significant new insight into the role that QCD plays
in the nucleon-nucleon interaction. DVCS and DVMP on polarized deuterons
will allow much of this rich structure to be extracted.

$^4$He is the ideal deeply-bound nucleus to start with,
since it is scalar and isoscalar, and thus admits a simple
description in terms of its spin and flavor structure.
In between this and $^2$H, $^3$He provides an opportunity
to study the $A$ dependence of nuclear effects, and
it could give easy access to neutron polarization properties,
due to its specific spin structure.
In addition, being isospin-$1/2$, it allows for the
flavor dependence of nuclear effects could be studied, 
in particular if parallel measurements 
on $^3$H targets were possible \cite{Scopetta:2009sn}.

Due to very small cross sections,
especially in the coherent channel where the nucleus does not break-up,
the addressed measurements are very difficult.
Despite this, the first data for coherent DVCS off $^4$He
collected at JLab in the 6~GeV era
have been published \cite{Hattawy:2017woc}.
A new impressive program is on the way at JLab,
carried on by the CLAS collaboration with the ALERT detector project 
\cite{Armstrong:2017zqr,Armstrong:2017zcm,Armstrong:2017wfw}.
Measurements for $^3$He and $^3$H
are not planned, but could be considered
as extensions of the ALERT project, at least in the unpolarized 
sector. Polarized measurements, which could
access neutron angular momentum information \cite{Rinaldi:2012pj,
Rinaldi:2012ft,Rinaldi:2014bba}, seem
very unlikely at JLab, due to the difficulty of arranging
a polarized target and a recoil detector in the same experimental setup. 

From the theoretical point of view, a realistic calculation 
of conventional effects for
nuclear few-body systems corresponds to a plane wave impulse approximation 
 analysis. This requires the evaluation of realistic non-diagonal spectral
functions for $^3$He and $^4$He.
For the first target, a complete analysis using the Av18
nucleon-nucleon (NN) potential is available
\cite{Scopetta:2009sn,Rinaldi:2012pj,
Rinaldi:2012ft,Rinaldi:2014bba,Scopetta:2004kj}.
Nuclear GPDs are found to be sensitive to details of 
the used NN interaction.
A study with nuclear ingredients of the same quality
for $^4$He is still missing and should be performed, to update 
existing calculations \cite{Guzey:2003jh,Liuti:2005gi}. 
The evaluation of a realistic spectral functions of $^4$He,
using state-of-the-art NN potentials,
will require the wave function of a three-body scattering state, 
which is a really challenging problem.
Besides, in the incoherent channel of DVCS off $^2$H, $^3$He, $^4$He,
even the study of specific final state interactions will be necessary.

An encouraging approximated calculation has been recently
performed for coherent DVCS off $^4$He \cite{PhysRevC.98.015203}, with the aim to
describe the CLAS data, as an intermediate
step towards a realistic evaluation.
A model of the nuclear non-diagonal spectral function, 
based on the momentum distribution
corresponding to the Av18 NN interaction
\cite{PhysRevC.67.034003}, has been used in the 
actual impulse approximation
calculation. Typical results
are found, in the proper limits, for the nuclear form factor
and for nuclear parton distributions.
Nuclear GPDs and the
Compton form factors are evaluated using
a well known GPD model to encompass the nucleonic information~\cite{Goloskokov:2011rd}. 
As can be seen in Fig.~\ref{alu}, a very good agreement 
is found with the data.  
One can conclude that 
a careful analysis of the reaction mechanism in terms of
basic conventional ingredients is successful and that
the present experimental accuracy does not require
the use of exotic arguments, such as dynamical off-shellness.
More refined nuclear calculations will be certainly necessary for 
the expected improved accuracy of the next generation of experiments 
at JLab, with the 12 GeV electron beam and high luminosity. 

Realistic calculations of $^2$H GPDs and Compton form factors using
state-of-the-art NN potentials are also available~\cite{Cano:2003ju}.
Due to the breaking of Lorentz covariance through a Fock space truncation,
the extraction of gravitational form factors---including those describing
the distributions of mass, spin, pressure, and shear---is ambiguous.
It is possible to model $^2$H GPDs using
a simpler contact interaction while maintaining covariance, and thus
make unambiguous predictions for GFFs and the stress-energy tensor.
This has the shortcoming of using a less realistic interaction,
and is limited to a domain where $t$ is very small. Manifestly covariant
calculations of nuclear structure with more realistic NN interactions
are sorely needed for GPD calculations, since GPDs are the only experimental
avenue through which GFFs can be accessed.
  
While JLab will provide important information
in the valence region, the extension to lower
$x$ values will be possible only at the EIC \cite{Accardi:2012qut}. 
Great opportunities at the EIC are also those related to the possibility
of easily using $^3$H beams, or polarized trinucleon beams and 
recoil detection at the same time, to access, for example, 
the neutron polarized structure and the flavor dependence 
of nuclear effects.

\section{Nuclear Dependence of $R=\sigma_L/\sigma_T$}
\label{sec:long}

Due to the relatively low energy of the 6~GeV JLab E03-103 measurements, effects due to acceleration and
deceleration of electrons in the Coulomb field of the heavier targets (Cu and Au) could not be ignored.
However, once applied, these so-called Coulomb corrections resulted in ratios systematically larger
than those found by previous experiments.  This apparent discrepancy motivated the re-examination of
earlier measurements, and it was found that, while the bulk of the large $x$ measurements from SLAC
were taken at significantly higher beam energies, SLAC E140 (an experiment dedicated to studying
the nuclear dependence of $R=\sigma_L/\sigma_T$, where $\sigma_L$ and $\sigma_T$ are the longitudinally and transversely polarized virtual photon component of the inelastic cross section) had used beam energies similar to JLab E03-103, but for
those measurements had not applied Coulomb corrections.  Once Coulomb corrections were applied to the E140
results, the possibility for concluding a nuclear dependence of $R$ was stronger but not unambiguous, with
$\Delta R=R_A-R_D$ about 1.5 standard deviations from zero. Further, when a combined analysis
of available SLAC E139, E140, and JLab E03-103 data was performed for data at $x=0.5$ and $Q^2\approx5$~GeV$^2$,
a similar deviation of $\Delta R$ was found~\cite{Solvignon:2009it}. Dedicated, high precision longitudinal/transverse separation experiments performed on hydrogen and deuterium at JLab also observed a likely nuclear dependence to $R=\sigma_L/\sigma_T$~\cite{Tvaskis:2016uxm,Tvaskis:2006tv}. Subsequent experiments on heavy nuclei in the resonance region at JLab, where effects have been predicted~\cite{Miller:2001yf} clearly demonstrated a nuclear dependence to $R${~\cite{Keppel_private} in this regime, although different from expected. 

The above observations and results, performed where $R$ is relatively large and therefore a nuclear dependence is measurable (as compared to previous high Q$^2$ measurements) motivated a 
new experiment to make further measurements of the nuclear dependence of $R=\sigma_L/\sigma_T$
with high precision and covering a wider range of $x$ and $Q^2$ than previously
measured~\cite{12gev_nucr}.  This experiment will provide precise measurements of $R=\sigma_L/\sigma_T$
for the nucleon for $0.1<x<0.6$ and $1<Q^2<5~\mathrm{GeV}^2$ as well as determination of $\Delta R= R_A-R_d$ for the same
kinematics using a copper target. Additional data will also be taken with C and Au targets for a subset
of the kinematics. Unanswered questions such as if the EMC effect is a longitudinal or transverse or combined effect will be answered by this measurement where, thus far, it has been incorrectly assumed that the longitudinal contribution is negligible. An analysis of the impact of the projected data points out that, in the presence of the observed small but non-zero difference between $R$ for nuclei and the nucleon, the nuclear enhancement in the ratio of the transverse structure functions $F_1^A/F_1^D$ becomes significantly reduced (or even disappears in some cases), indicating that anti-shadowing is dominated by the longitudinal contribution~\cite{Guzey:2012yk}. 

$R$, and correspondingly, the longitudinal nuclear inclusive structure function $F_L^A(x,Q^2)$, are quantities that directly probe the nuclear gluon distribution $g_A(x)$ in leading twist framework. The magnitude of nuclear enhancement to these structure functions is directly correlated with the size and shape of the nuclear gluons. While at the moment $g_A(x)$ is rather poorly constrained by QCD fits to available data, dedicated high-precision measurements of the nuclear dependence of $R$ at JLab and the EIC have the potential to constrain $g_A(x)$ in the anti-shadowing and EMC regions and beyond. Through the parton momentum sum rule, this knowledge will have some impact on $g_A(x)$ over the entire range of x, and so should also help to constrain $g_A(x)$ in the nuclear shadowing region, where the gluon distribution plays an essential role in the phenomenology of high-energy hard processes with nuclei.

\section{Systematics\label{sec:systematics}}

\subsection{Hadronization and Final State Effects}

The formation of hadrons and the propagation of quarks in a nuclear medium is important for interpreting the final states of reactions in semi-inclusive DIS and Drell-Yan processes as well as critical for interpreting gluon distributions.  An important open question are the relative sizes of the interactions between asymptotic quark propagation and interactions after hadronization.  As final states can reveal important information regarding the kinematics (such as Bjorken $x$) and flavor in a given interaction, improving the quality studies and models will simultaneously improve the extraction of modification data.  Studies have taken place at a variety of facilities, e.g. Fermilab~\cite{PhysRevC.75.035206}, JLab~\cite{PhysRevLett.99.242502, ELFASSI2012326}, and HERMES~\cite{Airapetian2011} as well as a future program with CLAS12 at JLab to study this in a broad set of channels~\cite{quarkformprop}.

Determination of the gluon distributions are also contaminated by final state interactions. nPDFs have been including routinely particle production in $d+\mathrm{Au}$ collisions into the fits, an observable that is very sensitive to both the initial state and hadronization of the gluon. The fragmentation functions for pion production have been best determined from $e^+e^-$ data, though the fragmentation of gluons have a sizable theoretical uncertainty. Data from the LHC at $7$~TeV can not be well described by a global fit unless a significant cut in $p_{T}$ is applied to the data, leaving out a relevant portion of the covered $d+\mathrm{Au}$ region.  With the present data, this constitutes a sizable source of uncertainty for the extraction of the nPDFs and conclusions from incorporating particle production data from hadron colliders into the fits must be drawn carefully.

\subsection{Free Nucleon Parton Distributions}

While the free proton parton distributions have been studied extensively and with great precision, there still remain important measurements to be done to constrain the two leading-flavor parton distributions, in particular in the ratio of $d/u$ limit as $x \rightarrow 1$.  As these represent the basis of comparison for any nuclear modification effect, it is critical to have high quality data available, especially as one considers doing flavor decompositions of nuclei.  There are several programs which intend to improve the fixed-target lepton scattering data, such as using the ratio of ${}^{3}$H and ${}^{3}$He cross section~\cite{mar}, tagged spectator with deuterium~\cite{bonus12}, and parity-violating deep inelastic scattering on the proton~\cite{solid_pvdis}.  In addition, recent analyses of $W$ and $Z$ production in $p\bar{p}$ collisions from the CDF and D\O\ collaborations~\cite{D0:2014kma,Abazov:2013dsa,Acosta:2005ud,Aaltonen:2009ta,Aaltonen:2010zza,Abazov:2007jy} have also provided new constraints at large $x$ and have been incorporated into a recent global PDF analysis~\cite{Accardi:2016qay}.

\section{Summary and Key Issues\label{sec:conclusion}}
The area of medium modification in nuclei has been a rich field for several decades and here we have provided a survey of the status and the outstanding issues to be addressed.  To continue to advance this field in a coherent manner, a world-wide effort is required on a broad number of topics.  We identify some of the most urgent questions along with their programs.
\begin{itemize}
\item What is the isovector nature of the EMC effect?  This requires a set of high-precision experiments across nuclei of traditional inclusive cross section ratios, electroweak measurements which are sensitive to unique quark flavor combinations, and Drell-Yan experiments.  These would be complemented by exclusive isotope tagging experiments.
\item What is the spin dependence of the EMC effect?  There is no experimental information available and any such measurement would break into new ground and would provide new information on potential mechanisms.
\item What is the momentum-dependence and virtuality-dependence of the EMC effect?  This can start to be addressed through tagged scattering measurements, for both low and high nucleon momenta, which can separate mean field and local density mechanisms, respectively.
\item What is the image of the full nucleus in terms of both quarks and gluons?  The study of generalized parton distributions through deeply virtual Compton scattering and deeply virtual meson production as well as at existing and future colliders are critical in producing a unified femtoscopic map of the nucleus.
\end{itemize}
Finally, the overarching question is how does the image of the nucleus in terms of quarks and gluons lead to a better understanding of the underlying nuclear dynamics and the origin of nuclei? Facilities such as Jefferson Lab, the proposed EIC, and other hadron physics facilities around the world we soon help us answer this important question.

\ack
The authors are grateful to the ECT* in Trento, Italy, for the hosting the workshop ``Exposing Novel Quark and Gluon Effects in Nuclei'' in April 2018 which made this work possible.  
The U.S. Department of Energy, Office of Science, Office of Nuclear Physics supported contributions from Argonne National Laboratory under contract DE-AC02-06CH11357. The contributions from the University of Tennesee, Knoxville, were supported by grant DE-SC0013615; contributions from the University of Washington were supported under Award Number DE-FG02-97ER-41014; and contributions from the University of Adelaide were supported by the Australian Research Council through the ARC Centre of Excellence for Particle Physics at the Terascale (CE110001104) and Discovery Project DP15110310.

\bibliography{main}

\providecommand{\newblock}{}
\begin{thebibliography}{100}
\expandafter\ifx\csname url\endcsname\relax
  \def\url#1{{\tt #1}}\fi
\expandafter\ifx\csname urlprefix\endcsname\relax\def\urlprefix{URL }\fi
\providecommand{\eprint}[2][]{\url{#2}}

\bibitem{PhysRevD.98.030001}
Tanabashi M {\em et~al.\/} (Particle Data Group) 2018 {\em Phys. Rev. D\/} {\bf
  98}(3) 030001
  \urlprefix\url{https://link.aps.org/doi/10.1103/PhysRevD.98.030001}

\bibitem{Taylor:1991ew}
Taylor R~E 1991 {\em Rev. Mod. Phys.\/} {\bf 63} 573--595

\bibitem{Benvenuti:1987az}
Benvenuti A~C {\em et~al.\/} (BCDMS) 1987 {\em Phys. Lett.\/} {\bf B189}
  483--487

\bibitem{Gomez:1993ri}
Gomez J {\em et~al.\/} 1994 {\em Phys. Rev.\/} {\bf D49} 4348--4372

\bibitem{Ashman:1992kv}
Ashman J {\em et~al.\/} (European Muon) 1993 {\em Z. Phys.\/} {\bf C57}
  211--218

\bibitem{Guzey:2012yk}
Guzey V, Zhu L, Keppel C~E, Christy M~E, Gaskell D, Solvignon P and Accardi A
  2012 {\em Phys. Rev.\/} {\bf C86} 045201 (\textit{Preprint}
  \eprint{1207.0131})

\bibitem{Aubert:1983xm}
Aubert J~J {\em et~al.\/} (European Muon) 1983 {\em Phys. Lett.\/} {\bf 123B}
  275--278

\bibitem{Altemus:1980wt}
Altemus R, Cafolla A, Day D, Mccarthy J, Whitney R {\em et~al.\/} 1980 {\em
  Phys. Rev. Lett.\/} {\bf 44} 965--968

\bibitem{Noble:1980my}
Noble J 1981 {\em Phys. Rev. Lett.\/} {\bf 46} 412--415

\bibitem{Meziani:1984is}
Meziani Z, Barreau P, Bernheim M, Morgenstern J, Turck-Chieze S {\em et~al.\/}
  1984 {\em Phys. Rev. Lett.\/} {\bf 52} 2130--2133

\bibitem{Morgenstern:2001jt}
Morgenstern J and Meziani Z 2001 {\em Phys. Lett.\/} {\bf B515} 269--275
  (\textit{Preprint} \eprint{nucl-ex/0105016})

\bibitem{Cloet:2015tha}
Cloët I~C, Bentz W and Thomas A~W 2016 {\em Phys. Rev. Lett.\/} {\bf 116}
  032701 (\textit{Preprint} \eprint{1506.05875})

\bibitem{Frankfurt:1988nt}
Frankfurt L~L and Strikman M~I 1988 {\em Phys. Rept.\/} {\bf 160} 235--427

\bibitem{Arneodo:1992wf}
Arneodo M 1994 {\em Phys. Rept.\/} {\bf 240} 301--393

\bibitem{Geesaman:1995yd}
Geesaman D~F, Saito K and Thomas A~W 1995 {\em Ann. Rev. Nucl. Part. Sci.\/}
  {\bf 45} 337--390

\bibitem{Piller:1999wx}
Piller G and Weise W 2000 {\em Phys. Rept.\/} {\bf 330} 1--94
  (\textit{Preprint} \eprint{hep-ph/9908230})

\bibitem{Sargsian:2002wc}
Sargsian M~M {\em et~al.\/} 2003 {\em J. Phys.\/} {\bf G29} R1--R45
  (\textit{Preprint} \eprint{nucl-th/0210025})

\bibitem{Malace:2014uea}
Malace S, Gaskell D, Higinbotham D~W and Cloet I 2014 {\em Int. J. Mod.
  Phys.\/} {\bf E23} 1430013 (\textit{Preprint} \eprint{1405.1270})

\bibitem{Hen:2016kwk}
Hen O, Miller G~A, Piasetzky E and Weinstein L~B 2017 {\em Rev. Mod. Phys.\/}
  {\bf 89} 045002 (\textit{Preprint} \eprint{1611.09748})

\bibitem{Seely:2009gt}
Seely J {\em et~al.\/} 2009 {\em Phys. Rev. Lett.\/} {\bf 103} 202301
  (\textit{Preprint} \eprint{0904.4448})

\bibitem{Norton:2003cb}
Norton P~R 2003 {\em Rept. Prog. Phys.\/} {\bf 66} 1253--1297

\bibitem{Hen:2013oha}
Hen O, Higinbotham D~W, Miller G~A, Piasetzky E and Weinstein L~B 2013 {\em
  Int. J. Mod. Phys.\/} {\bf E22} 1330017 (\textit{Preprint}
  \eprint{1304.2813})

\bibitem{Arrington:2012ax}
Arrington J, Daniel A, Day D, Fomin N, Gaskell D and Solvignon P 2012 {\em
  Phys. Rev.\/} {\bf C86} 065204 (\textit{Preprint} \eprint{1206.6343})

\bibitem{PhysRevC.82.054614}
Kulagin S~A and Petti R 2010 {\em Phys. Rev. C\/} {\bf 82}(5) 054614
  \urlprefix\url{https://link.aps.org/doi/10.1103/PhysRevC.82.054614}

\bibitem{Arrington:2003qt}
Arrington J 2004 {\em Acta Phys. Hung.\/} {\bf A21} 295--300 (\textit{Preprint}
  \eprint{hep-ph/0304213})

\bibitem{Mulders:1983au}
Mulders P~J and Thomas A~W 1984 {\em Phys. Rev. Lett.\/} {\bf 52} 1199

\bibitem{Frankfurt:1993sp}
Frankfurt L~L, Strikman M~I, Day D~B and Sargsian M 1993 {\em Phys. Rev.\/}
  {\bf C48} 2451--2461

\bibitem{Weinstein:2010rt}
Weinstein L~B, Piasetzky E, Higinbotham D~W, Gomez J, Hen O and Shneor R 2011
  {\em Phys. Rev. Lett.\/} {\bf 106} 052301 (\textit{Preprint}
  \eprint{1009.5666})

\bibitem{Hen:2012fm}
Hen O, Piasetzky E and Weinstein L~B 2012 {\em Phys. Rev.\/} {\bf C85} 047301
  (\textit{Preprint} \eprint{1202.3452})

\bibitem{PhysRevLett.119.262502}
Chen J~W, Detmold W, Lynn J~E and Schwenk A 2017 {\em Phys. Rev. Lett.\/} {\bf
  119}(26) 262502
  \urlprefix\url{https://link.aps.org/doi/10.1103/PhysRevLett.119.262502}

\bibitem{Fomin:2010ei}
Fomin N {\em et~al.\/} 2010 {\em Phys. Rev. Lett.\/} {\bf 105} 212502
  (\textit{Preprint} \eprint{1008.2713})

\bibitem{Tzanov:2005kr}
Tzanov M {\em et~al.\/} (NuTeV) 2006 {\em Phys. Rev.\/} {\bf D74} 012008
  (\textit{Preprint} \eprint{hep-ex/0509010})

\bibitem{Onengut:2005kv}
Onengut G {\em et~al.\/} (CHORUS) 2006 {\em Phys. Lett.\/} {\bf B632} 65--75

\bibitem{Tzanov:2009zz}
Tzanov M 2010 {\em AIP Conf. Proc.\/} {\bf 1222} 243--247

\bibitem{ATLAS:2014cpa}
Aad G {\em et~al.\/} (ATLAS) 2015 {\em Phys. Lett.\/} {\bf B748} 392--413
  (\textit{Preprint} \eprint{1412.4092})

\bibitem{Chatrchyan:2014hqa}
Chatrchyan S {\em et~al.\/} (CMS) 2014 {\em Eur. Phys. J.\/} {\bf C74} 2951
  (\textit{Preprint} \eprint{1401.4433})

\bibitem{Eskola:2016oht}
Eskola K~J, Paakkinen P, Paukkunen H and Salgado C~A 2017 {\em Eur. Phys. J.\/}
  {\bf C77} 163 (\textit{Preprint} \eprint{1612.05741})

\bibitem{PhysRevD.95.094013}
Klasen M, Kova\ifmmode~\check{r}\else \v{r}\fi{}\'{\i}k K and Potthoff J 2017
  {\em Phys. Rev. D\/} {\bf 95}(9) 094013
  \urlprefix\url{https://link.aps.org/doi/10.1103/PhysRevD.95.094013}

\bibitem{PhysRevD.97.114013}
Klasen M and Kova\ifmmode~\check{r}\else \v{r}\fi{}\'{\i}k K 2018 {\em Phys.
  Rev. D\/} {\bf 97}(11) 114013
  \urlprefix\url{https://link.aps.org/doi/10.1103/PhysRevD.97.114013}

\bibitem{PhysRevD.96.114005}
Aschenauer E~C, Fazio S, Lamont M~A~C, Paukkunen H and Zurita P 2017 {\em Phys.
  Rev. D\/} {\bf 96}(11) 114005
  \urlprefix\url{https://link.aps.org/doi/10.1103/PhysRevD.96.114005}

\bibitem{Egiyan:2003vg}
Egiyan K~S {\em et~al.\/} (CLAS) 2003 {\em Phys. Rev.\/} {\bf C68} 014313
  (\textit{Preprint} \eprint{nucl-ex/0301008})

\bibitem{Fomin:2011ng}
Fomin N {\em et~al.\/} 2012 {\em Phys. Rev. Lett.\/} {\bf 108} 092502
  (\textit{Preprint} \eprint{1107.3583})

\bibitem{12gev_emc}
Daniel A, Arrington J, Fomin N and Gaskell D 2010 {J}efferson {L}ab
  {E}xperiment {E}12-10-008
  \urlprefix\url{http://www.jlab.org/exp_prog/proposals/10/PR12-10-008.pdf}

\bibitem{12gev_xgt1}
Day D, Arrington J, Fomin N and Solvginon P 2006 {J}efferson {L}ab {E}xperiment
  {E}12-06-105
  \urlprefix\url{http://www.jlab.org/exp_prog/proposals/06/PR12-06-105.pdf}

\bibitem{mar}
Arrington J, Katramatou A~T, Petratos G~G and Ransome R~D 2010 {J}efferson
  {L}ab {E}xperiment {E}12-10-103
  \urlprefix\url{https://www.jlab.org/exp_prog/proposals/10/PR12-10-103.pdf}

\bibitem{tritsrc}
Hen O, Weinstein L, Gilad S and Boeglin W 2014 {J}efferson {L}ab {E}xperiment
  {E}12-14-011
  \urlprefix\url{https://www.jlab.org/exp_prog//proposals/14/PR12-14-011.pdf}

\bibitem{Londergan:2009kj}
Londergan J~T, Peng J~C and Thomas A~W 2010 {\em Rev. Mod. Phys.\/} {\bf 82}
  2009--2052 (\textit{Preprint} \eprint{0907.2352})

\bibitem{Cloet:2009qs}
Cloet I~C, Bentz W and Thomas A~W 2009 {\em Phys. Rev. Lett.\/} {\bf 102}
  252301 (\textit{Preprint} \eprint{0901.3559})

\bibitem{Cloet:2012td}
Cloet I~C, Bentz W and Thomas A~W 2012 {\em Phys. Rev. Lett.\/} {\bf 109}
  182301 (\textit{Preprint} \eprint{1202.6401})

\bibitem{Sargsian:2012sm}
Sargsian M~M 2014 {\em Phys. Rev.\/} {\bf C89} 034305 (\textit{Preprint}
  \eprint{1210.3280})

\bibitem{Arrington:2015wja}
Arrington J 2016 {\em EPJ Web Conf.\/} {\bf 113} 01011 (\textit{Preprint}
  \eprint{1508.05042})

\bibitem{Subedi:2008zz}
Subedi R {\em et~al.\/} 2008 {\em Science\/} {\bf 320} 1476--1478
  (\textit{Preprint} \eprint{0908.1514})

\bibitem{Zeller:2001hh}
Zeller G~P {\em et~al.\/} (NuTeV) 2002 {\em Phys. Rev. Lett.\/} {\bf 88} 091802
  [Erratum: Phys. Rev. Lett.90,239902(2003)] (\textit{Preprint}
  \eprint{hep-ex/0110059})

\bibitem{Paschos:1972kj}
Paschos E~A and Wolfenstein L 1973 {\em Phys. Rev.\/} {\bf D7} 91--95

\bibitem{Bentz:2009yy}
Bentz W, Cloët I~C, Londergan J~T and Thomas A~W 2010 {\em Phys. Lett.\/} {\bf
  B693} 462--466 (\textit{Preprint} \eprint{0908.3198})

\bibitem{Schienbein:2009kk}
Schienbein I, Yu J~Y, Kovarik K, Keppel C, Morfin J~G, Olness F and Owens J~F
  2009 {\em Phys. Rev.\/} {\bf D80} 094004 (\textit{Preprint}
  \eprint{0907.2357})

\bibitem{Patrignani:2016xqp}
Patrignani C {\em et~al.\/} (Particle Data Group) 2016 {\em Chin. Phys.\/} {\bf
  C40} 100001

\bibitem{emcpvdis}
Riordan S, Beminiwattha R and Arrington J 2016 {J}efferson {L}ab pac 44
  {P}roposal {PR}12-16-006
  \urlprefix\url{https://www.jlab.org/exp_prog/proposals/16/PR12-16-006.pdf}

\bibitem{Bodek:1983ec}
Bodek A {\em et~al.\/} 1983 {\em Phys. Rev. Lett.\/} {\bf 51} 534

\bibitem{Mineo:2003vc}
Mineo H, Bentz W, Ishii N, Thomas A~W and Yazaki K 2004 {\em Nucl. Phys.\/}
  {\bf A735} 482--514 (\textit{Preprint} \eprint{nucl-th/0312097})

\bibitem{Thomas:2016bxx}
Thomas A~W 2016 {\em EPJ Web Conf.\/} {\bf 123} 01003 (\textit{Preprint}
  \eprint{1606.05956})

\bibitem{Guichon:2018uew}
Guichon P~A~M, Stone J~R and Thomas A~W 2018 {\em Prog. Part. Nucl. Phys.\/}
  {\bf 100} 262--297 (\textit{Preprint} \eprint{1802.08368})

\bibitem{Stone:2017oqt}
Stone J, Guichon P and Thomas A 2017 {\em EPJ Web Conf.\/} {\bf 163} 00057
  (\textit{Preprint} \eprint{1706.01153})

\bibitem{Stone:2016qmi}
Stone J~R, Guichon P~A~M, Reinhard P~G and Thomas A~W 2016 {\em Phys. Rev.
  Lett.\/} {\bf 116} 092501 (\textit{Preprint} \eprint{1601.08131})

\bibitem{Cloet:2005rt}
Cloet I~C, Bentz W and Thomas A~W 2005 {\em Phys. Rev. Lett.\/} {\bf 95} 052302
  (\textit{Preprint} \eprint{nucl-th/0504019})

\bibitem{Cloet:2006bq}
Cloet I~C, Bentz W and Thomas A~W 2006 {\em Phys. Lett.\/} {\bf B642} 210--217
  (\textit{Preprint} \eprint{nucl-th/0605061})

\bibitem{jlabspin}
Brooks W~K and Kuhn S~E 2014 {J}efferson {L}ab {E}xperiment {E}12-14-001
  \urlprefix\url{https://www.jlab.org/exp_prog/proposals/14/PR12-14-001.pdf}

\bibitem{Tronchin:2018mvu}
Tronchin S, Matevosyan H~H and Thomas A~W 2018 {\em Phys. Lett.\/} {\bf B783}
  247--252 (\textit{Preprint} \eprint{1806.00481})

\bibitem{Sick:1992pw}
Sick I and Day D 1992 {\em Phys. Lett.\/} {\bf B274} 16--20

\bibitem{Thomas:1989vt}
Thomas A~W, Michels A, Schreiber A~W and Guichon P~A~M 1989 {\em Phys. Lett.\/}
  {\bf B233} 43--47

\bibitem{Guichon:1995ue}
Guichon P~A~M, Saito K, Rodionov E~N and Thomas A~W 1996 {\em Nucl. Phys.\/}
  {\bf A601} 349--379 (\textit{Preprint} \eprint{nucl-th/9509034})

\bibitem{Smith:2005ra}
Smith J~R and Miller G~A 2005 {\em Phys. Rev.\/} {\bf C72} 022203
  (\textit{Preprint} \eprint{nucl-th/0505048})

\bibitem{Thomas:2018kcx}
Thomas A~W 2018 {\em Int. J. Mod. Phys.\/} {\bf E27} 1840001 (\textit{Preprint}
  \eprint{1809.06622})

\bibitem{Pudliner:1997ck}
Pudliner B~S, Pandharipande V~R, Carlson J, Pieper S~C and Wiringa R~B 1997
  {\em Phys. Rev.\/} {\bf C56} 1720--1750 (\textit{Preprint}
  \eprint{nucl-th/9705009})

\bibitem{Bickerstaff:1985ax}
Bickerstaff R~P, Birse M~C and Miller G~A 1984 {\em Phys. Rev. Lett.\/} {\bf
  53} 2532--2535

\bibitem{Alde:1990im}
Alde D~M {\em et~al.\/} 1990 {\em Phys. Rev. Lett.\/} {\bf 64} 2479--2482

\bibitem{Vasilev:1999fa}
Vasilev M~A {\em et~al.\/} (NuSea) 1999 {\em Phys. Rev. Lett.\/} {\bf 83}
  2304--2307 (\textit{Preprint} \eprint{hep-ex/9906010})

\bibitem{seaquest}
Geesaman D~F, Reimer P~E {\em et~al.\/} 2006 {F}ermi {N}ational {A}ccelerator
  {L}aboratory {E}xperiment {E}906
  \urlprefix\url{http://www.phy.anl.gov/mep/SeaQuest/}

\bibitem{Badier:1981ci}
Badier J {\em et~al.\/} (NA3) 1981 {\em Phys. Lett.\/} {\bf 104B} 335
  [,807(1981)]

\bibitem{Bordalo:1987cs}
Bordalo P {\em et~al.\/} (NA10) 1987 {\em Phys. Lett.\/} {\bf B193} 368

\bibitem{Dutta:2010pg}
Dutta D, Peng J~C, Cloet I~C and Gaskell D 2011 {\em Phys. Rev.\/} {\bf C83}
  042201 (\textit{Preprint} \eprint{1007.3916})

\bibitem{Paakkinen:2016wxk}
Paakkinen P, Eskola K~J and Paukkunen H 2017 {\em Phys. Lett.\/} {\bf B768}
  7--11 (\textit{Preprint} \eprint{1609.07262})

\bibitem{Cazaroto:2008qh}
Cazaroto E~R, Carvalho F, Goncalves V~P and Navarra F~S 2008 {\em Phys.
  Lett.\/} {\bf B669} 331--336 (\textit{Preprint} \eprint{0804.2507})

\bibitem{Vogt:1999dw}
Vogt R 2000 {\em Phys. Rev.\/} {\bf C61} 035203 (\textit{Preprint}
  \eprint{hep-ph/9907317})

\bibitem{Alexandrov:1999ch}
Alexandrov {\relax Yu} {\em et~al.\/} (BEATRICE) 1999 {\em Nucl. Phys.\/} {\bf
  B557} 3--21

\bibitem{Piller:1995nc}
Piller G and Thomas A~W 1996 {\em Z. Phys.\/} {\bf C70} 661--664
  (\textit{Preprint} \eprint{hep-ph/9508410})

\bibitem{Modarres:2018ymh}
Modarres M and Hadian A 2018 {\em Phys. Rev.\/} {\bf D98} 076001
  (\textit{Preprint} \eprint{1901.06477})

\bibitem{Chang:2015qxa}
Chang E, Detmold W, Orginos K, Parreno A, Savage M~J, Tiburzi B~C and Beane S~R
  (NPLQCD) 2015 {\em Phys. Rev.\/} {\bf D92} 114502 (\textit{Preprint}
  \eprint{1506.05518})

\bibitem{Boeglin:2015cha}
Boeglin W and Sargsian M 2015 {\em Int. J. Mod. Phys.\/} {\bf E24} 1530003
  (\textit{Preprint} \eprint{1501.05377})

\bibitem{Cosyn:2017ekf}
Cosyn W and Sargsian M 2017 {\em Int. J. Mod. Phys.\/} {\bf E26} 1730004
  (\textit{Preprint} \eprint{1704.06117})

\bibitem{Frankfurt:1981mk}
Frankfurt L~L and Strikman M~I 1981 {\em Phys. Rept.\/} {\bf 76} 215--347

\bibitem{Keister:1991sb}
Keister B~D and Polyzou W~N 1991 {\em Adv. Nucl. Phys.\/} {\bf 20} 225--479
  [,225(1991)]

\bibitem{deutLDRD}
Weiss C {\em et~al.\/} 2014 {P}hysics potential of polarized light ions with
  EIC$@$JLab \urlprefix\url{https://www.jlab.org/theory/tag/}

\bibitem{Sargsian:2005rm}
Sargsian M and Strikman M 2006 {\em Phys. Lett.\/} {\bf B639} 223--231
  (\textit{Preprint} \eprint{hep-ph/0511054})

\bibitem{Klimenko:2005zz}
Klimenko A~V {\em et~al.\/} (CLAS) 2006 {\em Phys. Rev.\/} {\bf C73} 035212
  (\textit{Preprint} \eprint{nucl-ex/0510032})

\bibitem{Baillie:2011za}
Baillie N {\em et~al.\/} (CLAS) 2012 {\em Phys. Rev. Lett.\/} {\bf 108} 142001
  [Erratum: Phys. Rev. Lett.108,199902(2012)] (\textit{Preprint}
  \eprint{1110.2770})

\bibitem{Guzey:2014jva}
Guzey V, Higinbotham D, Hyde C, Nadel-Turonski P, Park K, Sargsian M, Strikman
  M and Weiss C 2014 {\em PoS\/} {\bf DIS2014} 234 (\textit{Preprint}
  \eprint{1407.3236})

\bibitem{Cosyn:2016oiq}
Cosyn W, Guzey V, Sargsian M, Strikman M and Weiss C 2016 {\em EPJ Web Conf.\/}
  {\bf 112} 01022 (\textit{Preprint} \eprint{1601.06665})

\bibitem{Strikman:2017koc}
Strikman M and Weiss C 2018 {\em Phys. Rev.\/} {\bf C97} 035209
  (\textit{Preprint} \eprint{1706.02244})

\bibitem{Hoodbhoy:1988am}
Hoodbhoy P, Jaffe R~L and Manohar A 1989 {\em Nucl. Phys.\/} {\bf B312}
  571--588

\bibitem{Airapetian:2005cb}
Airapetian A {\em et~al.\/} (HERMES) 2005 {\em Phys. Rev. Lett.\/} {\bf 95}
  242001 (\textit{Preprint} \eprint{hep-ex/0506018})

\bibitem{Slifer:2013vma}
Slifer K and Long E 2013 {\em PoS\/} {\bf PSTP2013} 008 (\textit{Preprint}
  \eprint{1311.4835})

\bibitem{Cosyn:2017fbo}
Cosyn W, Dong Y~B, Kumano S and Sargsian M 2017 {\em Phys. Rev.\/} {\bf D95}
  074036 (\textit{Preprint} \eprint{1702.05337})

\bibitem{Miller:2013hla}
Miller G~A 2014 {\em Phys. Rev.\/} {\bf C89} 045203 (\textit{Preprint}
  \eprint{1311.4561})

\bibitem{PhysRevC.95.014001}
Del~Dotto A, Pace E, Salm\`e G and Scopetta S 2017 {\em Phys. Rev. C\/} {\bf
  95}(1) 014001
  \urlprefix\url{https://link.aps.org/doi/10.1103/PhysRevC.95.014001}

\bibitem{Pace:2013bq}
Pace E, Salme G, Scopetta S, Del~Dotto A and Rinaldi M 2013 {\em Few Body
  Syst.\/} {\bf 54} 1079--1082 (\textit{Preprint} \eprint{1301.5787})

\bibitem{Scopetta:2014yoa}
Scopetta S, Del~Dotto A, Kaptari L, Pace E, Rinaldi M and Salmè G 2015 {\em
  Few Body Syst.\/} {\bf 56} 425--430 (\textit{Preprint} \eprint{1411.7559})

\bibitem{Pace:2016eiq}
Pace E, Del~Dotto A, Kaptari L, Rinaldi M, Salmé G and Scopetta S 2016 {\em
  Few Body Syst.\/} {\bf 57} 601--606 (\textit{Preprint} \eprint{1602.06521})

\bibitem{Dirac:1949cp}
Dirac P~A~M 1949 {\em Rev. Mod. Phys.\/} {\bf 21} 392--399

\bibitem{sid1}
Gao H, Chen J~P, Jiang X, Peng J~C and Qian X 2009 {J}efferson {L}ab {P}roposal
  {PR}12-09-014
  \urlprefix\url{https://www.jlab.org/exp_prog/proposals/09/PR12-09-014.pdf}

\bibitem{sid2}
Chen J~P, Qiang Y and Yan W 2011 {J}efferson {L}ab {P}roposal {PR}12-11-007
  \urlprefix\url{https://www.jlab.org/exp_prog/proposals/11/PR12-11-007.pdf}

\bibitem{Bakamjian:1953kh}
Bakamjian B and Thomas L~H 1953 {\em Phys. Rev.\/} {\bf 92} 1300--1310

\bibitem{Strauch:2002wu}
Strauch S {\em et~al.\/} (Jefferson Lab E93-049) 2003 {\em Phys. Rev. Lett.\/}
  {\bf 91} 052301 (\textit{Preprint} \eprint{nucl-ex/0211022})

\bibitem{Benhar:2006wy}
Benhar O, day D and Sick I 2008 {\em Rev. Mod. Phys.\/} {\bf 80} 189--224
  (\textit{Preprint} \eprint{nucl-ex/0603029})

\bibitem{CiofidegliAtti:1999kp}
Ciofi~degli Atti C, Kaptari L~P and Scopetta S 1999 {\em Eur. Phys. J.\/} {\bf
  A5} 191--207 (\textit{Preprint} \eprint{hep-ph/9904486})

\bibitem{CiofidegliAtti:2003pb}
Ciofi~degli Atti C, Kaptari L~P and Kopeliovich B~Z 2004 {\em Eur. Phys. J.\/}
  {\bf A19} 145--151 (\textit{Preprint} \eprint{nucl-th/0307052})

\bibitem{Alvioli:2006jd}
Alvioli M, Ciofi~degli Atti C and Palli V 2007 {\em Nucl. Phys.\/} {\bf A782}
  175--178 (\textit{Preprint} \eprint{nucl-th/0609030})

\bibitem{CiofidegliAtti:2007ork}
Ciofi~degli Atti C, Frankfurt L~L, Kaptari L~P and Strikman M~I 2007 {\em Phys.
  Rev.\/} {\bf C76} 055206 (\textit{Preprint} \eprint{0706.2937})

\bibitem{Armstrong:2017zqr}
Armstrong W {\em et~al.\/} 2017  (\textit{Preprint} \eprint{1708.00891})

\bibitem{Armstrong:2017zcm}
Armstrong W~R {\em et~al.\/} 2017  (\textit{Preprint} \eprint{1708.00835})

\bibitem{Dupre:2015jha}
Dupré R and Scopetta S 2016 {\em Eur. Phys. J.\/} {\bf A52} 159
  (\textit{Preprint} \eprint{1510.00794})

\bibitem{Berger:2001zb}
Berger E~R, Cano F, Diehl M and Pire B 2001 {\em Phys. Rev. Lett.\/} {\bf 87}
  142302 (\textit{Preprint} \eprint{hep-ph/0106192})

\bibitem{PhysRevC.98.015203}
Fucini S, Scopetta S and Viviani M 2018 {\em Phys. Rev. C\/} {\bf 98}(1) 015203
  \urlprefix\url{https://link.aps.org/doi/10.1103/PhysRevC.98.015203}

\bibitem{Hattawy:2017woc}
Hattawy M {\em et~al.\/} (CLAS) 2017 {\em Phys. Rev. Lett.\/} {\bf 119} 202004
  (\textit{Preprint} \eprint{1707.03361})

\bibitem{Scopetta:2009sn}
Scopetta S 2009 {\em Phys. Rev.\/} {\bf C79} 025207 (\textit{Preprint}
  \eprint{0901.3058})

\bibitem{Armstrong:2017wfw}
Armstrong W {\em et~al.\/} 2017  (\textit{Preprint} \eprint{1708.00888})

\bibitem{Rinaldi:2012pj}
Rinaldi M and Scopetta S 2012 {\em Phys. Rev.\/} {\bf C85} 062201
  (\textit{Preprint} \eprint{1204.0723})

\bibitem{Rinaldi:2012ft}
Rinaldi M and Scopetta S 2013 {\em Phys. Rev.\/} {\bf C87} 035208
  (\textit{Preprint} \eprint{1208.2831})

\bibitem{Rinaldi:2014bba}
Rinaldi M and Scopetta S 2014 {\em Few Body Syst.\/} {\bf 55} 861--864
  (\textit{Preprint} \eprint{1401.1350})

\bibitem{Scopetta:2004kj}
Scopetta S 2004 {\em Phys. Rev.\/} {\bf C70} 015205 (\textit{Preprint}
  \eprint{nucl-th/0404014})

\bibitem{Guzey:2003jh}
Guzey V and Strikman M 2003 {\em Phys. Rev.\/} {\bf C68} 015204
  (\textit{Preprint} \eprint{hep-ph/0301216})

\bibitem{Liuti:2005gi}
Liuti S and Taneja S~K 2005 {\em Phys. Rev.\/} {\bf C72} 032201
  (\textit{Preprint} \eprint{hep-ph/0505123})

\bibitem{PhysRevC.67.034003}
Viviani M, Kievsky A and Rinat A~S 2003 {\em Phys. Rev. C\/} {\bf 67}(3) 034003
  \urlprefix\url{https://link.aps.org/doi/10.1103/PhysRevC.67.034003}

\bibitem{Goloskokov:2011rd}
Goloskokov S~V and Kroll P 2011 {\em Eur. Phys. J.\/} {\bf A47} 112
  (\textit{Preprint} \eprint{1106.4897})

\bibitem{Cano:2003ju}
Cano F and Pire B 2004 {\em Eur. Phys. J.\/} {\bf A19} 423--438
  (\textit{Preprint} \eprint{hep-ph/0307231})

\bibitem{Accardi:2012qut}
Accardi A {\em et~al.\/} 2016 {\em Eur. Phys. J.\/} {\bf A52} 268
  (\textit{Preprint} \eprint{1212.1701})

\bibitem{Solvignon:2009it}
Solvignon P, Gaskell D and Arrington J 2009 {\em AIP Conf. Proc.\/} {\bf 1160}
  155--159 (\textit{Preprint} \eprint{0906.0512})

\bibitem{Tvaskis:2016uxm}
Tvaskis V {\em et~al.\/} 2018 {\em Phys. Rev.\/} {\bf C97} 045204
  (\textit{Preprint} \eprint{1606.02614})

\bibitem{Tvaskis:2006tv}
Tvaskis V {\em et~al.\/} 2007 {\em Phys. Rev. Lett.\/} {\bf 98} 142301
  (\textit{Preprint} \eprint{nucl-ex/0611023})

\bibitem{Miller:2001yf}
Miller G~A 2001 {\em Phys. Rev.\/} {\bf C64} 022201 (\textit{Preprint}
  \eprint{nucl-th/0104025})

\bibitem{Keppel_private}
Keppel C~E Private communication regarding results from Experiments E03-110 and
  E04-001 (publication in progress).

\bibitem{12gev_nucr}
Malace S, Christy M~E, Gaskell D, Keppel C~E and Solvignon P 2014 {J}efferson
  {L}ab {E}xperiment {E}12-14-002
  \urlprefix\url{http://www.jlab.org/exp_prog/proposals/14/PR12-14-002.pdf}

\bibitem{PhysRevC.75.035206}
Johnson M~B, Kopeliovich B~Z, Leitch M~J, McGaughey P~L, Moss J~M, Potashnikova
  I~K and Schmidt I 2007 {\em Phys. Rev. C\/} {\bf 75}(3) 035206
  \urlprefix\url{https://link.aps.org/doi/10.1103/PhysRevC.75.035206}

\bibitem{PhysRevLett.99.242502}
Clasie B, Qian X, Arrington J {\em et~al.\/} 2007 {\em Phys. Rev. Lett.\/} {\bf
  99}(24) 242502
  \urlprefix\url{https://link.aps.org/doi/10.1103/PhysRevLett.99.242502}

\bibitem{ELFASSI2012326}
Fassi L~E, Zana L, Hafidi K, Holtrop M, Mustapha B, Brooks W, Hakobyan H, Zheng
  X {\em et~al.\/} 2012 {\em Physics Letters B\/} {\bf 712} 326 -- 330 ISSN
  0370-2693
  \urlprefix\url{http://www.sciencedirect.com/science/article/pii/S0370269312005369}

\bibitem{Airapetian2011}
{The HERMES Collaboration}, Airapetian A, Akopov N, Akopov Z, Aschenauer E~C,
  Augustyniak W, Avakian R, Avetissian A, Avetisyan E, Belostotski S, Bianchi
  N, Blok H~P, Borissov A, Bowles J, Brodski I, Bryzgalov V {\em et~al.\/} 2011
  {\em The European Physical Journal A\/} {\bf 47} 113 ISSN 1434-601X
  \urlprefix\url{https://doi.org/10.1140/epja/i2011-11113-5}

\bibitem{quarkformprop}
Hafidi K {\em et~al.\/} 2006 {J}efferson {L}ab {P}roposal {PR}12-06-117
  \urlprefix\url{http://www.jlab.org/exp_prog/proposals/06/PR12-06-117.pdf}

\bibitem{bonus12}
Bueltmann S, Christy M~E, Howard F, Griffioen K~A, Keppel C~E, Kuhn S~E,
  Melnitchouk W and Tvaskis V 2006 {J}efferson {L}ab {P}roposal {PR}12-06-113
  \urlprefix\url{https://www.jlab.org/exp_prog/proposals/06/PR12-06-113.pdf}

\bibitem{solid_pvdis}
Souder P {\em et~al.\/} 2010 {J}efferson {L}ab {P}roposal {PR}12-10-007
  \urlprefix\url{https://www.jlab.org/exp_prog/proposals/10/PR12-10-007.pdf}

\bibitem{D0:2014kma}
Abazov V~M {\em et~al.\/} (D0) 2015 {\em Phys. Rev.\/} {\bf D91} 032007
  [Erratum: Phys. Rev.D91,no.7,079901(2015)] (\textit{Preprint}
  \eprint{1412.2862})

\bibitem{Abazov:2013dsa}
Abazov V~M {\em et~al.\/} (D0) 2014 {\em Phys. Rev. Lett.\/} {\bf 112} 151803
  [Erratum: Phys. Rev. Lett.114,no.4,049901(2015)] (\textit{Preprint}
  \eprint{1312.2895})

\bibitem{Acosta:2005ud}
Acosta D {\em et~al.\/} (CDF) 2005 {\em Phys. Rev.\/} {\bf D71} 051104
  (\textit{Preprint} \eprint{hep-ex/0501023})

\bibitem{Aaltonen:2009ta}
Aaltonen T {\em et~al.\/} (CDF) 2009 {\em Phys. Rev. Lett.\/} {\bf 102} 181801
  (\textit{Preprint} \eprint{0901.2169})

\bibitem{Aaltonen:2010zza}
Aaltonen T~A {\em et~al.\/} (CDF) 2010 {\em Phys. Lett.\/} {\bf B692} 232--239
  (\textit{Preprint} \eprint{0908.3914})

\bibitem{Abazov:2007jy}
Abazov V~M {\em et~al.\/} (D0) 2007 {\em Phys. Rev.\/} {\bf D76} 012003
  (\textit{Preprint} \eprint{hep-ex/0702025})

\bibitem{Accardi:2016qay}
Accardi A, Brady L~T, Melnitchouk W, Owens J~F and Sato N 2016 {\em Phys.
  Rev.\/} {\bf D93} 114017 (\textit{Preprint} \eprint{1602.03154})

\end{thebibliography}
\bibliographystyle{iopart-num}

\end{document}